\documentclass[prb,twocolumn,superscriptaddress,showpacs,floatfix]{revtex4}
\usepackage{mathrsfs}
\usepackage{amsmath,amsfonts,amssymb}
\usepackage{epsfig}
\usepackage{graphicx}
\usepackage{wasysym}
\usepackage{bbm}
\usepackage{psfrag}
\usepackage{color}
\usepackage{pstool}
\usepackage{braket}
\usepackage{lmodern}
\usepackage{ulem}
\usepackage[dvips, bookmarks, colorlinks=true, plainpages = false, citecolor = blue, linkcolor = blue, urlcolor = blue, filecolor = blue]{hyperref}
\def\be{\begin{equation}}
\def\ee{\end{equation}}
\def\bea{\begin{eqnarray}}
\def\eea{\end{eqnarray}}

\begin{document}
\title{Quantum phases of spin-1/2  extended XY model in transverse magnetic field}
\author{Rakesh Kumar Malakar}\email{rkmalakar75@gmail.com}
\author{Asim Kumar Ghosh}
 \email{asimkumar96@yahoo.com}
\affiliation {Department of Physics, Jadavpur University, 
188 Raja Subodh Chandra Mallik Road, Kolkata 700032, India}
\begin{abstract}
  In this study, a spin-1/2 extended anisotropic XY chain has been introduced
  in which both time reversal and SU(2) symmetries are broken
  but $Z_2$ symmetry is preserved. 
Magnetic and topological phase diagrams in the parameter space 
have been drawn in the presence of transverse magnetic field. 
Entanglement measures like
  mutual information and quantum discord are also evaluated and it indicates that
  these transitions are second order in nature. 
  Quantum phase transition is noted
  at zero magnetic field, as well as  magnetic long range order is found to
  withstand magnetic field of any strength. 
  Exact analytic results for spin-spin correlation functions have been
  obtained in terms of Jordan Wigner fermionization.
  Existence of long range magnetic order has been investigated 
  numerically by finding correlation functions as well as the Binder cumulant
  in the ground state. 
  Dispersion relation, ground state energy, and energy gap are obtained analytically.
  In order to find the topologically nontrivial phase, sign of Pfaffian invariant and 
  value of winding number have  been evaluated. Both magnetic and topological phases are robust
  against the magnetic field and found to move coercively in the parameter space
  with the variation of its strength. Long range orders along
  two orthogonal directions and two different topological phases are found 
  and their one-to-one correspondence has been found. 
  Finally casting the spinless fermions onto Majorana fermions, properties of
  zero energy edge states are studied. Three different kinds of Majorana pairings are
  noted. In the trivial phase, next-nearest-neighbor Majorana pairing
  is found, whereas two different types of
  nearest-neighbor Majorana pairings are identified 
  in the topological superconducting phase.
  \vskip 1 cm
  Corresponding author: Asim Kumar Ghosh,

  email: asimkumar96@yahoo.com
\end{abstract}
\maketitle
\section{INTRODUCTION}
Quantum fluctuation driven physical phenomena are so widespread and exotic
that they have constituted several branches of study within the 
condensed matter physics. Among them the most important branch is
quantum phase transition (QPT)\cite{Sachdev,BKC1,BKC3,Franchini}.
QPT deals with the transition between phases of different types of quantum 
orders driven solely by quantum fluctuations as they
occur at zero-temperature ($T\!=\!0$). Nowadays, there have been an upsurge
of investigations on the topological orders of quantum systems and
transition between different topological phases.
So, it is pertinent to understand the 
interplay between QPTs and topological transitions as well as their
possible interconnections. The most studied quantum systems which exhibit
QPTs are spin-1/2 one-dimensional (1D) Ising and anisotropic XY 
models under the transverse magnetic field
\cite{LSM,Katsura,Barouch1,Barouch2,Barouch3,Pfeuty}. 
Exact analytic form of spin-spin correlation functions and other
quantities had been obtained by converting spin operators
to spinless fermions by Jordan-Wigner (JW) transformations \cite{JW}. 
Those systems exhibit magnetic long-range-order (LRO) and undergoes
transition to disorder phase 
at a specific value of magnetic field,
where spin fluctuations play crucial role for
the QPTs. Signature of QPT has been observed in ferromagnet 
CoNb$_2$O$_6$, ferroelectric KH$_2$PO$_4$, and other quantum matters
\cite{Coldea,Blinc,Stinchcombe}.

Quantum fluctuations have led to another important
phenomenon which is known as entanglement that 
exhibits unusual feature in the vicinity of QPT point. For examples,
derivatives of concurrences for anisotropic XY chain in
transverse field show logarithmic divergences at the
transition points\cite{Osterloh,Osborne}.
Entanglement in the ground state of the anisotropic XY chain
has been evaluated in terms of von Neumann and Renyi entropies
\cite{Korepin1,Korepin2,Franchini1}. 
The constant entropy surfaces are either ellipse or
hyperbolas and they are found to meet at the QPT points. 
Entanglement has been recognized as the key feature for the development of
quantum computation and communication\cite{Nielsen}.

Investigation on topological matters have begun with the
discovery of Quantum Hall Effect by K Von Klitzing\cite{Klitzing}. 
In this phenomenon, Hall conductivity of two-dimensional
(2D) electron gas got quantized under strong magnetic field,
which was subsequently characterized by the topological invariant known as 
Thouless-Kohmoto-Nightingale-Nijs (TKNN) number\cite{TKNN}.
This branch of study has been enriched further with the discovery of 
quantum anomalous Hall effect, where quantization of Hall Conductivity 
was made possible only with the time reversal symmetry (TRS) breaking
complex hopping terms, replacing the true magnetic field\cite{Haldane}.
Quantum matters exhibiting this particular quantized feature are termed as topological
insulator. This phenomenon can be understood easily with the help of
paradigmatic Su-Schrieffer-Heeger (SSH) model which is nothing but a 1D 
tight-binding system composed of two-site unit cells with staggered
hopping parameters\cite{SSH1,SSH2,Rakesh}.
The nontrivial phase is characterized by 
a nonzero topological invariant known as
winding number which is determined by integrating the 
Berry curvature over the 1D Brillouin zone (BZ). 
Band gap must be nonzero for the nontrivial phase which
is accompanied additionally 
by the symmetry protected zero energy boundary states, those are found 
localized on the edges of the open chain. Number of edge states in a
particular phase is 
determined by the bulk-boundary correspondence rule\cite{Hatsugai}. 
However, at the topological transition point band gap must vanish. 

The same topological order has been achieved 
separately through another investigation in terms of 
1D Kitaev model, which has opened up another branch of
study known as topological superconductivity\cite{Kitaev}.
Actually transverse-field XY 
model assumes the form of Kitaev model under
JW transformation where the faithful coexistence of
magnetic and topological superconducting
phases has been established\cite{Franchini}. 
In this case, band gap of the system 
can be identified specifically as the superconducting gap
for Cooper pairing of parallel spin. 
After expressing fermionic operators in terms of Majorana 
fermions, different types of Majorana pairings are found to
emerge in the trivial and topological phases.
Topological matter may provide higher efficiency
in electronic transport due to the fact that
electron motion in the nontrivial phase is
immune from the back scattering caused by the 
defects or disorders present in the materials. 
Therefore, better understanding on the role of quantum fluctuations
is required where both long-range quantum orders and topological phases
are found to emerge simultaneously and coexist.  

In order to study the effect of spin fluctuations for topological
matter, a 1D spin-1/2 model
composed of the anisotropic XY chain and a three spin term
has been introduced in this work, and its 
property has been investigated extensively.
The system has been solved exactly by expressing the spin operators
in terms of JW fermions where analytic forms of spin-spin correlations
functions have been obtained. In addition, magnetic property of the system
has been studied numerically by estimating the ground state correlations 
along with the Binder cumulant of the magnetic LRO for
a finite chain. 
Topological phases of the system have been characterized
analytically by obtaining the eigenvectors of quasiparticle dispersion spectra.
Magnetic phase is detected in the ground state, while the
topological phase is found in the single particle
spinless fermionic excitation.
System exhibits two distinct magnetic LROs
as well as two different topologically nontrivial
superconducting orders and they are found to coexist
exhibiting the one-to-one correspondence between
them.  
Comprehensive phase diagrams for magnetic and topological phases
are drawn. Transition between different magnetic phases is also associated to 
that between respective topological phases.
Peculiarities of these phases attribute to the fact that they
can withstand the magnetic field of any strength while  
transitions between them will occur even without the magnetic field.


The article has been arranged in the following order. 
Hamiltonian is introduced in the 
Sec \ref{model}, as well as symmetries of the model 
have been described. Various physical quantities are
defined in order to study the character of magnetic LRO.
Properties of the Hamiltonian at several limits have been
studied in great details and
compared with the known results. 
Magnetic properties of the system have been
investigated in the Sec \ref{magnet}, which has begun with the
analytic results for four-spin plaquette, followed by numerical
results obtained by exact diagonalization and analytic results
by JW fermionization. Dispersion relation,
band gap, spin-spin correlation functions, Binder cumulant and
magnetic phase diagram have been obtained.  
Topological properties of the system have been presented
in Sec \ref{topology}. Values of Pfaffian invariant and winding number for
different topological phases have been estimated here.
Edge states are determined and their properties in
terms of Majorana pairing have been studied. 
Topological phase diagram has been drawn.
Properties of the system in terms of entanglement measures like
mutual information and quantum discord have been presented in the
section \ref{Entanglement}. Finally, 
an extensive discussion based on all the results has been made available 
in Sec \ref{Discussion}.
\section{Anisotropic extended XY chain in transverse magnetic field}
\label{model}
Hamiltonian (Eq. \ref{ham}) of the 1D anisotropic extended XY model in the
presence of transverse magnetic field is written as
\bea
   H&=&\sum_{j=1}^N\big[J((1+\gamma)S_j^xS_{j+1}^x
  +(1-\gamma)S_j^yS_{j+1}^y) \nonumber\\[-0.2em]
    &&\;\;+J'(S_{j}^xS_{j+2}^x + S_{j}^yS_{j+2}^y)S_{j+1}^z+h_{\rm z}S_j^z\big],
 \label{ham}
  \eea
where, $N$ is the total number of sites and
$J$ is the nearest neighbor (NN) exchange interaction strength. 
$\gamma$ is the anisotropic parameter while
$J'$ is the three-spin exchange interaction strength.
$S_j^\alpha,\,\alpha=x,y,z$, is the $\alpha$-component of
spin-1/2 operator at
the site $j$, and $h_{\rm z}$ is the strength of magnetic field
acting along the $z$ direction.
This particular three-spin term 
can be labelled as `XZX+YZY' type of interaction. 
Hamiltonian breaks the 
rotational symmetry, $U(1)$, about any direction,
since $[H,\,S_{\rm T}^\alpha]\ne 0$, where
$S_{\rm T}^\alpha$ is the $\alpha$-component of
the total spin, $\boldsymbol S_{\rm T}$. 
The TRS is broken due to the
presence of odd-spin terms, which means that the Hamiltonian
retains TRS when $J'=0$, and $h_{\rm z}=0$. 
Symmetries of $H$ in the spin space under five different
transformations and their consequences 
are studied extensively which are discussed below. 
\subsection{Symmetries of $H$:}
Hamiltonian obeys the symmetry relation,
\be U_z H(J,J',\gamma,h_{\rm z})\, U^\dag_z =
H(-J,J',\gamma,h_{\rm z}),
\label{Sym1}\ee
where $ U_z=\prod_{j=1}^Ne^{i\pi jS_j^z}$, and when
 $N$ is assumed even.
The operator $U_z$ rotates the spin vector at the $j$-th site
about the $z$-axis by the angle $j\pi$.
This symmetry claims that energy spectrum of $H$ 
must remain unchanged upon sign inversion of $J$.
In the same way, it can be shown that $H$ satisfies the relation,
\be V_\beta H(J,J',\gamma,h_{\rm z}) V_\beta^\dag=H(J,-J',\gamma,-h_{\rm z}),
\label{Sym2}\ee
where $V_\beta=\prod_{j=1}^Ne^{i\pi S_j^\beta},\,\beta=x,\, {\rm or}\,y$. 
$V_\beta$ performs rotation of the every spin vector about
the $\beta$-axis by the angle $\pi$. It signifies that
energy spectrum is unaltered upon simultaneous sign inversion of
both $J'$ and $h_{\rm z}$.
As a consequence, Hamiltonian 
remains invariant under the combined operations, $W=V_\beta U_z$, since 
\be W H(J,J',\gamma,h_{\rm z}) W^\dag=-H(J,J',\gamma,h_{\rm z}).
\label{Sym3}\ee
The relation, $W H W^\dag=-H,$ corresponds to the
fact that the energy spectrum
of $H$ inherits the inversion symmetry around the zero energy,
or, it reflects the particle-hole like symmetry of the system. 
It further implies that zero energy states must appear in pair
if the spectrum possesses them. 
This feature has a special importance in the context of
nontrivial topological phase where the emergence of 
pair of zero-energy edge states in the 1D open system corresponds to 
the nonzero value of bulk topological invariant\cite{Hatsugai}.

Under the rotation of each spin vector about the $z$-axis by
the angle $\pi/2$, which is accomplished by the
operator $R_z=\prod_{j=1}^Ne^{i\frac{\pi}{2} S_j^z}$,  
$H$ undergoes the transformation like 
\be R_z H(J,J',\gamma,h_{\rm z}) R^\dag_z =
H(J, J',-\gamma,h_{\rm z}).
\label{Sym4}\ee
It reveals that the energy spectrum of $H$ remains invariant
under sign reversal of $\gamma$. 
In other words, it implies that as long as $\gamma \ne 0$,
$H$ lacks the $U(1)$ symmetry. 
However, $H$ remains invariant finally under the transformation of
$V_z=\prod_{j=1}^Ne^{i\pi S_j^z}$, which means
\be V_z H(J,J',\gamma,h_{\rm z}) V_z^\dag=H(J,J',\gamma,h_{\rm z}).
\label{Sym5}\ee
As $V_z$ performs rotation of the every spin vector about
the $z$-axis by the angle $\pi$,
the Hamiltonian possesses the $Z_2$  symmetry.
Effect of these symmetries on the properties of
the system will be discussed in the proper context. 
Magnetic and topological properties of $H$ will be
described in the subsequent sections. Antiferromagnetic (AFM) phase will appear
when $J$ is assumed positive, and that will be replaced by ferromagnetic (FM)
phase when $J$ is turned negative. In the next section,
several operators for identifying magnetic orders 
employed in numerical or analytic approaches will be discussed.
\subsection{Long range correlations:}
Existence of long range correlation of magnetic order in a system 
can be studied by evaluating either the
(i) uniform correlation functions
 ${\mathcal C}_{\rm FM}^\alpha (n),\,\alpha=x,y,z$, 
for the FM order or  
(ii) Neel (staggered spin-spin) correlation functions
 ${\mathcal C}_{\textrm{N\'eel}}^\alpha (n)$, 
for the AFM order.
The expressions of these functions are given as 
\bea
{\mathcal C}_{\rm FM}^\alpha (n)&=&\langle{\mathcal O}_{\rm FM}^\alpha (n)\rangle,
\; n=0,1,2,\cdots (N\!-\!1), \,{\rm where,}\nonumber \\[0.5em]
   {\mathcal O}_{\rm FM}^\alpha (n)&=&\frac{1}{N}\sum_{j=1}^N S_j^\alpha \,S_{j+n}^\alpha,
  \;{\rm and},\nonumber \\
{\mathcal C}_{\textrm{N\'eel}}^\alpha (n)&=&\langle{\mathcal O}_{\textrm{N\'eel}}^\alpha (n)\rangle,
\; \,{\rm where,}\nonumber \\[0.5em]
   {\mathcal O}_{\textrm{N\'eel}}^\alpha (n)&=&\frac{1}{N}\sum_{j=1}^N(-1)^n  S_j^\alpha \,S_{j+n}^\alpha,
\label{CF}
\eea
where ${\mathcal O}_{\rm FM/{\textrm {N\'eel}}}^\alpha (n)$ is the operator of FM/AFM order.
The number $n$ denotes the separation between two spins
between which the correlation is to be measured. 
In order to mark the phase transition points,
Binder cumulant for FM/AFM orders\cite{Binder1,Binder2}:
\be{\mathcal B}_{\rm FM/{\textrm {N\'eel}}}^\alpha=1-\frac{\langle (M_{\rm FM/{\textrm {N\'eel}}}^{^\alpha })^4\rangle}
   {3\;\langle (M_{\rm FM/{\textrm {N\'eel}}}^{^\alpha })^2\rangle^2},
\label{BP}
\ee   
can be evaluated numerically for the chains of finite length, where 
\be M_{\rm FM}^\alpha
=\frac{1}{N}\sum_{j=1}^N S_j^\alpha,\;\;M_{\textrm{N\'eel}}^\alpha
=\frac{1}{N}\sum_{j=1}^N (-1)^j S_j^\alpha,
\label{Mag}
\ee
are the operators for uniform and staggered magnetizations,
respectively.
In every case, expectation value $\langle * \rangle$ has been 
evaluated with respect to the ground state.
One can further check that 
\be (M_{\rm FM/{\textrm {N\'eel}}}^{^\alpha})^2=N^2\,\sum_{n=0}^{N-1}{\mathcal O}_{\rm FM/{\textrm {N\'eel}}}^\alpha (n).
\label{MtoC}
\ee
Definition of ${\mathcal B}_{\rm FM/{\textrm {N\'eel}}}^\alpha$ (Eq. \ref{BP})
indicates that the maximum possible value of it is
2/3. This maximum value will be assumed when the system
develops the perfect LRO.  
Correlation functions for $H$ obey the relations:
${\mathcal C}_{\rm FM/{\textrm {N\'eel}}}^x(n)\ne{\mathcal C}_{\rm FM/{\textrm {N\'eel}}}^y (n)
\ne{\mathcal C}_{\rm FM/{\textrm {N\'eel}}}^z (n)$ as long as $\gamma \ne 0$.
But ${\mathcal C}_{\rm FM/{\textrm {N\'eel}}}^x(n)={\mathcal C}_{\rm FM/{\textrm {N\'eel}}}^y (n)$,
when $\gamma=0$, as $U(1)$ symmetry is being preserved
by the system in this case. 
Similar relation also holds for the Binder cumulants,
${\mathcal B}_{\rm FM/{\textrm {N\'eel}}}^\alpha$.
However, in this study, Binder cumulant has been evaluated
to establish the existence of AFM LRO only. 
\subsection{Limiting cases of $H$:}
At several limits of the Hamiltonian, system is found identical
to well known spin models whose characteristics
have been studied extensively long before.
Here, properties of such five different models are described briefly. 
\subsubsection{For $\gamma=\pm 1$, $J'=0$, and $h_{\rm z}=0$.}
Hamiltonian reduces to Ising models, however the spin operators
align along the directions orthogonal to each other, since 
\[H_{\rm x}=2J\sum_{j=1}^NS_j^xS_{j+1}^x,\;{\rm and},\; 
H_{\rm y}=2J\sum_{j=1}^NS_j^yS_{j+1}^y,\]
respectively, when $\gamma=\pm 1$\cite{Ising}.
These Hamiltonians are connected to each other via
rotation of spin operators about the $z$-axis by
$\pm \pi/2$. Apart from this those two
Hamiltonians are equivalent to each other
in every aspect for obvious reasons like they
possess LRO but exhibit no QPT. 
The ground state is doubly degenerate for each Hamiltonian in which 
$Z_2$ symmetry is broken and the system possesses a gap. 
At those points, Hamiltonians retain their
rotational symmetry about either $x$ and $y$ directions, where
AFM (FM) LRO is favoured along the respective directions
as long as $J>0$ ($J<0$). 
As a result, ground states correspond to a pair of N\'eel states
in terms of tensor product of eigenspinors of  $S^x$ and $S^y$ operators.
So, in order to construct the exact ground state wave functions 
at the degenerate point $\gamma=1$, for the thermodynamic limit, 
the normalized eigenspinors of $S^x$ operator are defined
as $S^x|\chi^\pm\rangle=\pm \frac{1}{2}|\chi^\pm\rangle$.
Eigenspinors can be further expressed as
$|\chi^\pm\rangle=\frac{1}{\sqrt 2}(|\!\uparrow\rangle\pm |\!\downarrow\rangle)$,
where $S^z|\!\!\uparrow\rangle=+\frac{1}{2}|\!\!\uparrow\rangle$, and
 $S^z|\!\downarrow\rangle=-\frac{1}{2}|\!\downarrow\rangle$.
At this point of the parameter space,
Hamiltonian, $H_{\rm x}$ is found to commute with the
staggered spin-spin operator, 
${\mathcal O}_{\textrm {N\'eel}}^\alpha (n)$, for $\alpha=x$,
but not for  $\alpha=y$ and $z$. 
The exact form of doubly degenerate
normalized ground states at this point is given by the product of
single spin states as 
\be |\Psi^\pm_{\rm x}\rangle=\bigotimes_{m=1,2,3,\cdots}^{N/2}\left(|\chi^\pm\rangle_{2m-1}
\otimes |\chi^\mp\rangle_{2m} \right),
\label{Gstate-x}\ee
where total number of sites, $N$ is even. Obviously, the pair of
ground states satisfies the orthonormality condition,
$\langle\Psi^\mu_{\rm x}|\Psi^\nu_{\rm x} \rangle=\delta_{\mu\nu}$,
where $(\mu,\nu)=\pm$. Further, they are
connected to each other by lattice translation of unity. 

Finally, it can be shown that ground state,
$|\Psi^\pm_{\rm x}\rangle$ satisfies the
eigen value equations,
\be \left\{\begin{array}{c}
H_{\rm x}|\Psi^\pm_{\rm x}\rangle=-\frac{J}{2}N|\Psi^\pm_{\rm x}\rangle,\\[0.9em]
 {\mathcal O}_{\textrm {N\'eel}}^x (n)|\Psi^\pm_{\rm x}\rangle=\frac{1}{4}|\Psi^\pm_{\rm x}\rangle,
\end{array}\right.
\label{eigenvalues-x}\ee
where the eigenvalues actually correspond to the exact ground state
energy per site, $E_{\rm G}=-\frac{J}{2}$, and the exact value of
the $x$-component for staggered correlation functions,
${\mathcal C}_{\textrm {N\'eel}}^x (n)=1/4$, which is independent
of $n$. At the same time,
they yield
\[{\mathcal C}_{\textrm {N\'eel}}^\alpha (n)=\langle\Psi^\pm_{\rm x}|
  {\mathcal O}_{\textrm {N\'eel}}^\alpha (n)|\Psi^\pm_{\rm x}\rangle=0,\]
when  $\alpha=y$ and $z$. It means AFM LRO of the
$x$-component of spin-spin correlation exists and no LRO persists for the
$y$ and $z$ components of that. 
It is worth mentioning at that point that no other
types of correlations like FM and
chiral orders are found here. However, AFM phase will
be replaced by the FM one if $J<0$, with the same value of
exact ground state energy per site, $E_{\rm G}=-\frac{|J|}{2}$,
and the exact value of $x$-component for spin-spin correlation functions,
${\mathcal C}_{\rm FM}^x (n)=1/4$. 

Similarly, at another degenerate point, say, $\gamma=-1$, 
$J'=0$, and $h_{\rm z}=0$, ground state of $H_{\rm y}$, 
{\it i. e.}, $\Psi^\pm_{\rm y}$ can be constructed in terms of 
eigenspinors of $S^y$ operator as shown below: 
\be |\Psi^\pm_{\rm y}\rangle=\bigotimes_{m=1,2,3,\cdots}^{N/2}\left(|\eta^\pm\rangle_{2m-1}
\otimes |\eta^\mp\rangle_{2m} \right),
\label{Gstate-y}\ee
where $S^y|\eta^\pm\rangle=\pm \frac{1}{2}|\eta^\pm\rangle$, and 
$|\eta^\pm\rangle=\frac{1}{\sqrt 2}(|\!\uparrow\rangle\pm i\, |\!\downarrow\rangle)$.  
Obviously, 
$|\Psi^\pm_{\rm y}\rangle$ satisfies the
eigen value equations,
\be \left\{\begin{array}{c}
H_{\rm y}|\Psi^\pm_{\rm y}\rangle=-\frac{J}{2}N|\Psi^\pm_{\rm y}\rangle,\\[0.9em]
 {\mathcal O}_{\textrm {N\'eel}}^y (n)|\Psi^\pm_{\rm y}\rangle=\frac{1}{4}|\Psi^\pm_{\rm y}\rangle,
\end{array}\right.
\label{eigenvalues-y}\ee
So, again $E_{\rm G}=-\frac{J}{2}$, and 
the $y$-component for spin-spin staggered correlation functions,
${\mathcal C}_{\textrm {N\'eel}}^y (n)=1/4$. 
At the same time,
they yield the relations, 
\[{\mathcal C}_{\textrm {N\'eel}}^\alpha (n)=\langle\Psi^\pm_{\rm y}|
  {\mathcal O}_{\textrm {N\'eel}}^\alpha (n)|\Psi^\pm_{\rm y}\rangle=0,\]
when  $\alpha=x$ and $z$ for obvious reasons. 
So, both the points $\gamma = \pm 1$, system
exhibits AFM LRO at temperature $T=0$, as long as $J>0$.
FM phase will replace the AFM phase whenever $J<0$.
 The doubly degenerate ground states in the FM phase
can be written in terms of the tensor product of eigenspinors for 
 $S^x_m$ and $S^y_m$ operators as\cite{Korepin2}
\be |\Psi^\pm_{\rm x}\rangle=\!\!\!\bigotimes_{m=1,2,3,\cdots}^{N}\!\!\!|\chi^\pm\rangle_{m},\quad
 |\Psi^\pm_{\rm y}\rangle=\!\!\!\bigotimes_{m=1,2,3,\cdots}^{N}\!\!\!|\eta^\pm\rangle_{m},
\label{Gstate-FM}\ee
replacing the respective Eqs. \ref{Gstate-x} and \ref{Gstate-y}, in the AFM case 
for $\gamma=\pm 1$. Ground state having this kind of product structure
is extremely rare for a system with three-spin interations. 

Interestingly enough, the system at
those points ($\gamma=\pm 1$) exhibits nontrivial
topological characteristics in addition both for
FM and AFM phases\cite{Franchini}.
This feature can be shown
when the single particle
states are studied in terms of spinless fermions
by means of JW fermionization. 
 Under JW transformation, system assumes nothing but the
form of $p$-wave superconductor, which is
known as the isotropic Kitaev chain with zero
chemical potential \cite{Franchini,Kitaev}.
It is isotropic in a sense that
value of hopping parameter becomes equal to that of superconducting
parameter, and that is equal to NN exchange strength $J$.
In terms of this parametrization, the system always exhibit the
nontrivial topological phase with $\nu=\pm 1$, for
$\gamma=\pm 1$, since the chemical potential is absent. 
\subsubsection{For $\gamma=\pm 1$, $J'=0$, and $h_{\rm z}\ne 0$.}
When $h_{\rm z}\ne 0$, Hamiltonian
turns into the transverse Ising model, where
the system hosts LRO as long as $h_{\rm z}/J<1$, for $T=0$.
The system undergoes a phase transition at $h_{\rm z}/J=1$,
to the disordered phase for $h_{\rm z}/J \ge 1$, in
which ground state
preserves the $Z_2$ symmetry of the Hamiltonian\cite{Sachdev,BKC1,BKC3,Pfeuty}.
The next-nearest-neighbor (NNN) concurrence for $T=0$
exhibits a sharp peak at the transition point\cite{Osborne}. 
 Since $h_{\rm z}$ plays the role of chemical potential whenever the 
Hamiltonian expressed in JW fermions, the resulting model becomes equal to isotropic
Kitaev chain with nonzero chemical potential. 
As the values of hopping and superconducting parameters are equal,
system suffers a topological phase transition at the point 
$h_{\rm z}/J=1$, separating the nontrivial topological phase ($\nu=1$)
for $h_{\rm z}/J<1$, from the trivial phase ($\nu=0$) for  $h_{\rm z}/J\ge 1$.
So, the results clearly show that topological and magnetic phases
coexist for both the cases, $h_{\rm z}= 0$, and $h_{\rm z}\ne 0$.
\subsubsection{For $\gamma=0$, $J'=0$, and $h_{\rm z}\ne 0$.}
At this isotropic point, Hamiltonian preserves the $U(1)$ symmetry. 
Correlation functions for $h_{\rm z}= 0$, behave as\cite{Barouch2} 
\[{\mathcal C}_{\textrm {N\'eel}}^\beta (n)=\left\{\begin{array}{cc}
(-1)^{\frac{n+1}{2}}\,\frac{2}{n\pi},
&{\rm if}\;n\in{\rm odd},\\[0.8em]
0,&{\rm if}\;n\in{\rm even},\end{array}\right.\]
where $\beta=x,y$. The system exhibits 
short range order and so hosts no magnetic LRO, since
\[\lim_{n\rightarrow \infty}{\mathcal C}_{\textrm {N\'eel}}^\beta (n)=0.\]
Also there is no question for 
LRO for arbitrary $h_{\rm z}$. 
Energy gap vanishes at this point and it separates
two ordered phases around it.   
Phase transition that occurs across the line $\gamma=0$
is known as anisotropy transition, 
where the correlation $\mathcal C_{\textrm{N\'eel}}^x(n)$ 
($\mathcal C_{\textrm{N\'eel}}^y(n)$) for arbitrary $n$ 
survives when $\gamma>0$ ($\gamma<0$), though the
magnetization $M_{\textrm{N\'eel}}^x$ ($M_{\textrm{N\'eel}}^y$) vanishes. 
The system exhibits second-order phase transition
at  $h_z/J=\pm 1$, where disordered spin polarized phases
appear when  $h_z/J> 1$ and $h_z/J<-1$\cite{Katsura}. 
As a result, system hosts two 
bicritical points each one at $\gamma=0$, and $h_z/J=\pm 1$\cite{BKC3}. 
 However, no topological phase exists in this case
for any value of $h_{\rm z}$. 
\subsubsection{For $\gamma \ne 0$, $J'=0$, and $h_{\rm z}\ne 0$.}
Under this condition, the model is
known as anisotropic XY model in a transverse magnetic field.
The resulting spin model is exactly solvable, and there is
an energy gap. 
For $\gamma>0$, the asymptotic value of the correlation functions leads to \cite{Barouch2} 
\be\lim_{n\rightarrow \infty}{\mathcal C}_{\textrm {N\'eel}}^x (n)\!=\!\left\{\begin{array}{cc}
\!\!\frac{1}{2(1+\gamma)}\left[\gamma^2\left\{1\!-\!\left(\frac{h_{\rm z}}{J}\right)^2\right\} \right]^{1/4}\!\!,
&{\rm if}\;|h_{\rm z}|<J,\\[1.1em]
0,&{\rm if}\;|h_{\rm z}|\ge J,\end{array}\right.
\label{asymptotic}
\ee
and 
\[\lim_{n\rightarrow \infty}{\mathcal C}_{\textrm {N\'eel}}^y (n)=0.\]
With the sign reversal of $\gamma$, correlation function,
${\mathcal C}_{\textrm {N\'eel}}^x (n)$, will be replaced by
${\mathcal C}_{\textrm {N\'eel}}^y (n)$. 
The magnetic LRO exists within the range, $-1<h_{\rm z}/J<1$, irrespective of
the value of $\gamma$. 
 System hosts the topological phase with $\nu= + 1$ ($\nu=-1$), for $\gamma>0$
($\gamma<0$).  
Obviously, the asymptotic behavior of correlation functions described 
in the cases {\it  1, 2, 3} and {\it  4} can be derived from the
more general expressions of them in Eq. \ref{asymptotic}.

  Further, ground state is found doubly degenerate
over the circle
\be \left(\frac{h_{\rm z}}{J}\right)^2+\gamma^2=1, 
\label{circle} \ee
drawn in the parameter space. For $\gamma \ge 0$, the normalized ground
states can be expressed as the product of single spin states
in terms of the angular parameter $\theta$ ($-\pi/2\le \theta \le \pi/2$):  
\be |\Psi^\pm_{\rm \theta}\rangle=\bigotimes_{m=1,2,3,\cdots}^{N/2}\left(|\theta^\pm\rangle_{2m-1}
\otimes |\theta^\mp\rangle_{2m} \right),
\label{Gstate-theta}\ee
where  the single spin states are given by 
\(|\theta^\pm\rangle=(\cos{\theta}\,|\!\uparrow\rangle\pm
\sin{\theta}\,|\!\downarrow\rangle)\)\cite{Franchini,Korepin2,Kurmann1,Kurmann2,Muller}.
For the fixed $\gamma$, value of
$\theta$ could be determined by the roots of
quadratic equation: $\cos^2{(2\theta)}=(1-\gamma)/(1+\gamma)$.
For the Ising point, $(\gamma,h_{\rm z})=(1,0)$,
the roots are $\theta_\pm=\pm \pi/4$, which correspond to the
doubly degenerate ground states given in Eq \ref{Gstate-x}.
At the other limiting points given by $(\gamma,h_{\rm z})=(0,\pm 1)$,
the roots are $\theta_+=0$, and $\theta_-=\pi/2$, respectively, 
which correspond to two different 
states polarized along two opposite directions of $h_{\rm z}$,
but having the same energy. Ground states for the FM case are given by\cite{Franchini}
\[ |\Psi^\pm_{\rm \theta}\rangle=\bigotimes_{m=1,2,3,\cdots}^{N}|\theta^\pm\rangle_{m}.\]
Remarkably, asymptotic behavior of the correlations,
${\mathcal C}_{\textrm {N\'eel}}^\beta(n\rightarrow \infty)$,
is found different around this circle.
They contain oscillatory terms within the circle,
while outside the circle they are monotonic\cite{Barouch2}.
Interplay of magnetic correlation and entanglement in this case
seems much interesting by noting that von Neumann entropy has minimum over
the circle (Eq \ref{circle})\cite{Korepin1}. 
\subsubsection{For $\gamma=0$, and $J'\ne 0$.}
Disperson relation and magnetic properties of the system 
under this condition 
have been studied before in terms of
spinless fermion \cite{Japaridze}.
The system exhibits no long-range magnetic order in the 
absence of magnetic field, however, 
it undergoes transition between
two different spin-liquid phases when
$J'/J=2$.
The ground state phase
diagram in the presence of both uniform and staggered
magnetic field are obtained, where spin-polarized FM and AFM 
phases are found to appear, respectively, when the
strength of the field is very high.
True magnetic LRO has been developed in a system in order to
minimize the cooperative exchange energy, while the
spin-polarized phase appears when the Zeeman energy
overcomes the exchange energy.
In this study FM version of the model
was considered since $J<0$\cite{Japaridze}. 
Ground state phase diagram with different types of
spin-liquid phases have been described.  
 The system is no longer topological for any values of $J'/J$, and
$h_{\rm z}$ in this case. 

Summarizing the results described in the last five cases
it indicates that no LRO exists when $\gamma=0$, as
shown in the cases {\it 3} and {\it 5}. Magnetic LRO
exits when $\gamma\ne 0$ as shown in the cases
{\it 1, 2,} and {\it 4}, but at the same time it indicates that AFM 
${\mathcal C}_{\textrm {N\'eel}}^x$ survives for $J>0$, when
$\gamma>0$, while ${\mathcal C}_{\textrm {N\'eel}}^y$ survives
for $\gamma<0$. In the same way, topological phases with $\nu=1$
exits when $\gamma>0$, while that with $\nu=-1$ appears for $\gamma<0$.
 It means LRO with ${\mathcal C}_{\textrm {N\'eel}}^x$
(${\mathcal C}_{\textrm {N\'eel}}^y$)
corresponds to topological phase with $\nu=1$ ($\nu=-1$).
So one-to-one correspondence between magnetic and topological
phases has been shown. 
The above argument holds but AFM correlations,
${\mathcal C}_{\textrm {N\'eel}}^x$ and ${\mathcal C}_{\textrm {N\'eel}}^y$
would be replaced by FM correlations,
${\mathcal C}_{\textrm {FM}}^x$ and ${\mathcal C}_{\textrm {FM}}^y$, 
in the respective cases when $J<0$. 
The same correspondence
is still valid in the presence of three-spin terms which
is being established in the present work.
\subsubsection{For $\gamma \ne 0$, $J'\ne 0$, and $h_{\rm z}\ne 0$.}
The most general case has been studied extensively in this work. 
It is still exactly solvable by means of spinless fermion. 
Both magnetic LRO and topological phases are found present for
this model and they coexist in the parameter space.
It means that the non-zero value of $\gamma$ simultaneously
induces both magnetic and topological orders.
Phase diagrams for both the orders have been made.
The effect of three-spin interacting term has indicated that
unlike the previous cases, magnetic field is
not only incapable to destroy the LRO but also no more indispensable
for QPTs. As a result,
parameter regime exhibiting LRO is not bounded
within a limited parameter space as long as $h_{\rm z}/J'$
is finite, since with the variation of $h_{\rm z}$,
region holding LRO is found to shift its location 
but without altering its extent. 
Topological phases are detected as usual 
by evaluating the Pfaffian invariant, winding number and 
symmetry protected zero-energy edge states. 
For this purpose, the low energy single particle
dispersion relation is obtained in terms of
JW fermionization. In order to find magnetic LRO, 
spin-spin correlation functions,
${\mathcal C}_{\textrm{N\'eel}}^\beta (n)$ have been evaluated.
Additionally, AFM LRO has been
detected numerically by finding the value of them 
for the system with finite number of sites using
Lanczos exact diagonalization. Precise boundary of AFM phase is found by
evaluating the Binder cumulant in addition.
In the subsequent sections magnetic and topological properties of
this proposed model will be discussed and it begins
with the analytic results for four-spin plaquette of $H$
as described below.
\section{Magnetic properties of $H$:}
\label{magnet}
\subsection {Four-spin plaquette for $H$:}
The four-site Hamiltonian ($N=4$) under the periodic boundary condition
(PBC) can be obtained analytically. The exact expression for 
all the 16 eigen values ($E_m, m=1,2,\cdots, 16$)
and eigen functions ($\psi_m$) of $H$ are shown in Appendix
\ref{appendix:eigensystem}.
Energy spectrum consists of two pairs of
zero energy states. This feature is consistent with the symmetry of the
Hamiltonian, although they are not related to topological
edge modes by any means. 
$\psi_1$ is the ground state 
as long as $h_{\rm z}=0$ and $\gamma=1$
for any values of $J'$. In addition, 
$\psi_5$ is found degenerate with $\psi_1$
only when $J'=0$.
However, no ground state crossover is there
for $h_{\rm z}=0$ and $\gamma=1$, but 
ground state is doubly degenerate when $J'=0$. 
This does not hold for $\gamma \ne |1|$.
Whereas, for $h_{\rm z} \ne 0$, 
$\psi_1$ is the ground state
in the region, $-J'_{\rm c}<J'<J'_{\rm c}$, 
where
\[J'_{\rm c}=2\sqrt{\left(\sqrt{\eta+\sqrt{\eta^2-8h_z^2J^2}}-J\right)^2
  -J^2\gamma^2}-2h_z,\] 
and $\eta=J^2(1+\gamma^2)+2h_z^2$. $\psi_5$ is the ground state
beyond this region in this case.
It means ground state crossover is there, however, this
occurrence cannot be related to the phase
transition anymore as it attributes to finite size effect.
\begin{figure}[h]
  \psfrag{Eg}{$E_{\rm G}/J$}
   \psfrag{C-1}{ ${\mathcal C}_{\textrm{N\'eel}}^x(1)$}
\psfrag{a}{(a)}
\psfrag{b}{(b)}
\psfrag{c}{(c)}
\psfrag{d1}{$\gamma\!=\!1$}
\psfrag{h0}{$h_{\rm z}\!=\!0$}
\psfrag{h1}{$h_{\rm z}\!=\!1$}
\psfrag{h-1}{$h_{\rm z}\!=\!-1$}
\psfrag{Jp}{$J'/J$}
\psfrag{0}{0}
\psfrag{1}{1}
\psfrag{2}{2}
\psfrag{-1}{$-1$}
\psfrag{-3}{$-3$}
\psfrag{-2}{$-2$}
\psfrag{-4}{$-4$}
\psfrag{3}{3}
\psfrag{4}{4}
\psfrag{-0.5}{$-0.5$}
\psfrag{-0.6}{$-0.6$}
\psfrag{-0.7}{$-0.7$}
\psfrag{-0.8}{$-0.9$}
\psfrag{-0.9}{$-0.9$}
\psfrag{0.15}{$0.15$}
\psfrag{0.20}{$0.20$}
\psfrag{0.25}{$0.25$}
\includegraphics[width=230pt]{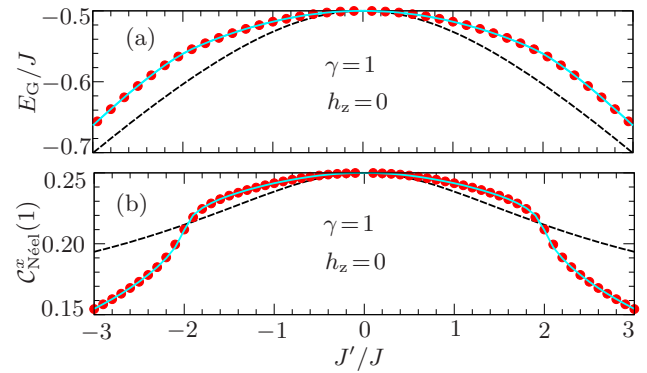}
\caption{Variation of ground state energy per site, $E_{\rm G}$ 
  (a), and correlation function,
 ${\mathcal C}_{\textrm{N\'eel}}^x(1)$ (b), for $-3\le J'/J \le 3$,
  when $\gamma=1$, and $h_{\rm z}=0$.
  Red circles are the numerical data while black dashed line for four-spin plaquette.
  Exact result, Eq. \ref{Eq-Eg}, in (a) and Eq. \ref{JWCx}, 
  in (b) are plotted in cyan line for comparison. Analytic and numerical
results show an excellent agreement.}
 \label{ground-state-energy-correlation}
\end{figure}

\begin{figure}[h]
  \psfrag{CNx}{ ${\mathcal C}_{\textrm{N\'eel}}^x(n)$}
   \psfrag{n}{\large $n$}
    \psfrag{a}{(a)}
    \psfrag{b}{(b)}
  \psfrag{c}{(c)}
  \psfrag{d}{(d)}
  \psfrag{e}{(e)}
  \psfrag{d1}{$\gamma=1$}
  \psfrag{jp}{ $J'/J$}
   \psfrag{x}{ $2/3$}
  \psfrag{N=28}{$N\!\!=\!\!28$}
\psfrag{N=20}{$N\!\!=\!\!20$}
\psfrag{N=24}{$N\!\!=\!\!24$}
\psfrag{N=12}{$N\!\!=\!\!12$}
\psfrag{N=16}{$N\!\!=\!\!16$}
\psfrag{h=0}{$h_{\rm z}\!=\!0$}
\psfrag{h=1}{$h_{\rm z}\!=\!1$}
\psfrag{h=2}{$h_{\rm z}\!=\!2$}
\psfrag{h=-1}{$h_{\rm z}\!=\!-1$}
\psfrag{h=-2}{$h_{\rm z}\!=\!-2$}
\psfrag{0}{0}
\psfrag{1}{1}
\psfrag{2}{2}
\psfrag{3}{3}
\psfrag{4}{4}
\psfrag{5}{5}
\psfrag{6}{6}
\psfrag{7}{7}
\psfrag{8}{8}
\psfrag{10}{10}
\psfrag{12}{12}
\psfrag{14}{14}
\psfrag{16}{16}
\psfrag{20}{20}
\psfrag{24}{24}
\psfrag{28}{28}
\psfrag{-1}{$-1$}
\psfrag{-3}{$-3$}
\psfrag{-2}{$-2$}
\psfrag{-4}{$-4$}
\psfrag{-5}{$-5$}
\psfrag{-6}{$-6$}
\psfrag{-7}{$-7$}
\psfrag{0.1}{$0.1$}
\psfrag{0.2}{$0.2$}
\psfrag{0.3}{$0.3$}
\psfrag{0.4}{$0.4$}
\psfrag{0.0}{$0.0$}
\psfrag{0.5}{$0.5$}
\psfrag{0.6}{$0.6$}
\psfrag{0.7}{$0.7$}
\psfrag{0.00}{$0.00$}
\psfrag{0.05}{$0.05$}
\psfrag{0.10}{$0.10$}
\psfrag{0.15}{$0.15$}
\psfrag{0.20}{$0.20$}
\psfrag{0.25}{$0.25$}
\includegraphics[width=230pt]{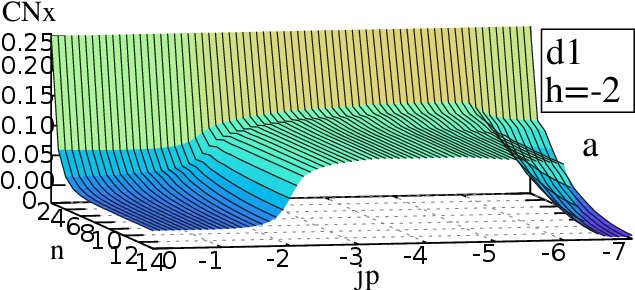}
\includegraphics[width=230pt]{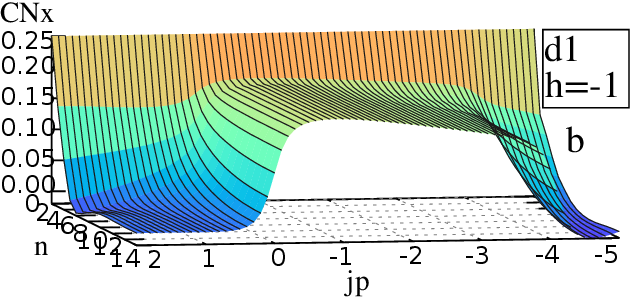}
\includegraphics[width=230pt]{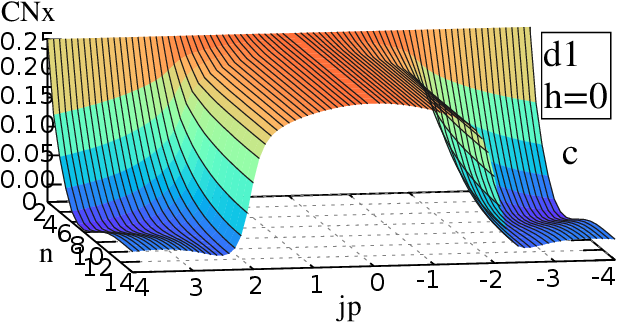}
\includegraphics[width=230pt]{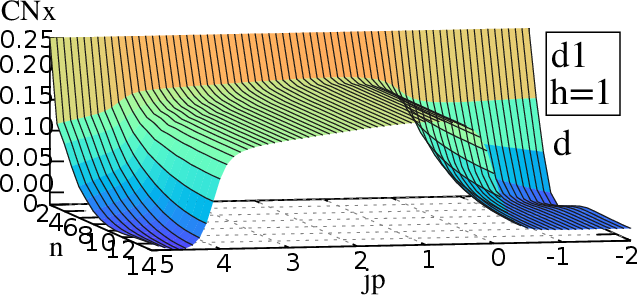}
\includegraphics[width=230pt]{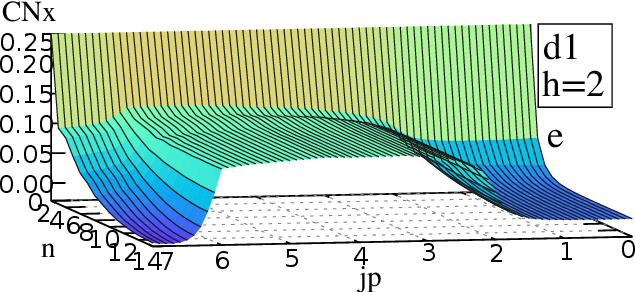}
\caption{Correlation, ${\mathcal C}_{\textrm{N\'eel}}^x(n)$, 
  when $\gamma=1$, for $h_{\rm z}=-2$ (a), $h_{\rm z}=-1$ (b), $h_{\rm z}=0$ (c),
 $h_{\rm z}=1$ (d), and $h_{\rm z}=2$ (e).}
 \label{Correlation-Neelx-1}
\end{figure}
\begin{figure}[h]
  \psfrag{CNx}{ ${\mathcal C}_{\textrm{N\'eel}}^x(n)$}
   \psfrag{n}{\large $n$}
    \psfrag{a}{(a)}
    \psfrag{b}{(b)}
  \psfrag{c}{(c)}
  \psfrag{d}{(d)}
  \psfrag{e}{(e)}
  \psfrag{d2}{$\gamma=2$}
  \psfrag{jp}{ $J'/J$}
   \psfrag{x}{ $2/3$}
  \psfrag{N=28}{$N\!\!=\!\!28$}
\psfrag{N=20}{$N\!\!=\!\!20$}
\psfrag{N=24}{$N\!\!=\!\!24$}
\psfrag{N=12}{$N\!\!=\!\!12$}
\psfrag{N=16}{$N\!\!=\!\!16$}
\psfrag{h=0}{$h_{\rm z}\!=\!0$}
\psfrag{h=1}{$h_{\rm z}\!=\!1$}
\psfrag{h=2}{$h_{\rm z}\!=\!2$}
\psfrag{h=-1}{$h_{\rm z}\!=\!-1$}
\psfrag{h=-2}{$h_{\rm z}\!=\!-2$}
\psfrag{0}{0}
\psfrag{1}{1}
\psfrag{2}{2}
\psfrag{3}{3}
\psfrag{4}{4}
\psfrag{5}{5}
\psfrag{6}{6}
\psfrag{7}{7}
\psfrag{8}{8}
\psfrag{10}{10}
\psfrag{12}{12}
\psfrag{14}{14}
\psfrag{16}{16}
\psfrag{20}{20}
\psfrag{24}{24}
\psfrag{28}{28}
\psfrag{-1}{$-1$}
\psfrag{-3}{$-3$}
\psfrag{-2}{$-2$}
\psfrag{-4}{$-4$}
\psfrag{-5}{$-5$}
\psfrag{-6}{$-6$}
\psfrag{-7}{$-7$}
\psfrag{0.1}{$0.1$}
\psfrag{0.2}{$0.2$}
\psfrag{0.3}{$0.3$}
\psfrag{0.4}{$0.4$}
\psfrag{0.0}{$0.0$}
\psfrag{0.5}{$0.5$}
\psfrag{0.6}{$0.6$}
\psfrag{0.7}{$0.7$}
\psfrag{0.00}{$0.00$}
\psfrag{0.05}{$0.05$}
\psfrag{0.10}{$0.10$}
\psfrag{0.15}{$0.15$}
\psfrag{0.20}{$0.20$}
\psfrag{0.25}{$0.25$}
\includegraphics[width=230pt]{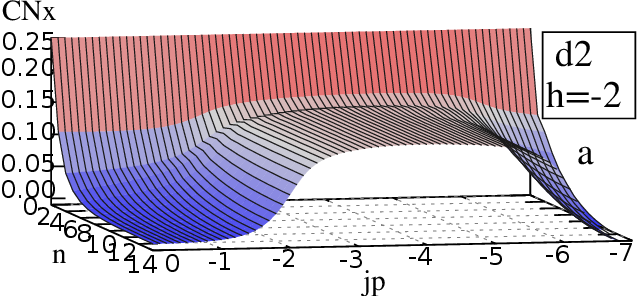}
\includegraphics[width=230pt]{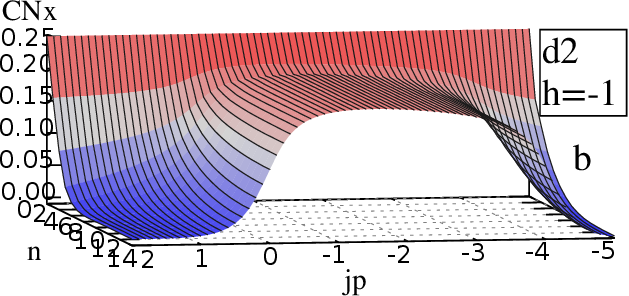}
\includegraphics[width=230pt]{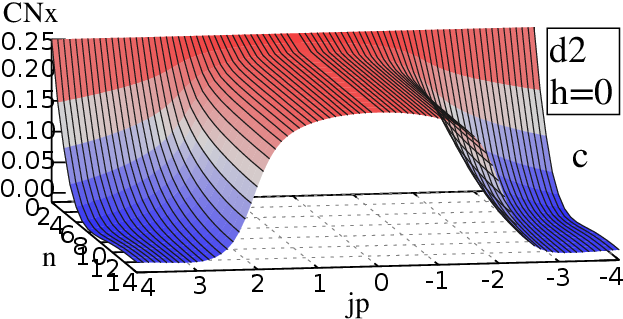}
\includegraphics[width=230pt]{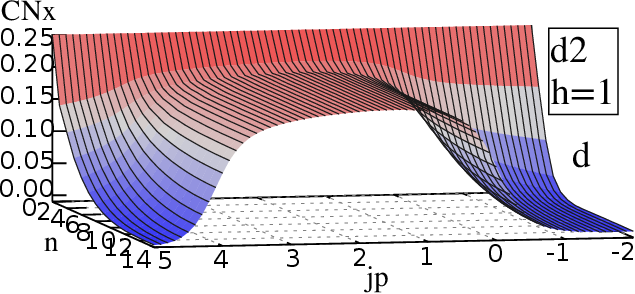}
\includegraphics[width=230pt]{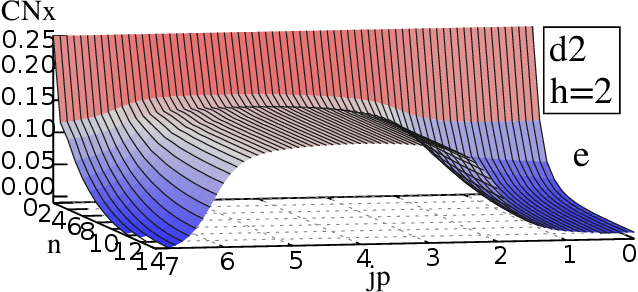}
\caption{Correlation, ${\mathcal C}_{\textrm{N\'eel}}^x(n)$,
  when $\gamma=2$, for $h_{\rm z}=-2$ (a), $h_{\rm z}=-1$ (b), $h_{\rm z}=0$ (c),
 $h_{\rm z}=1$ (d), and $h_{\rm z}=2$ (e).}
 \label{Correlation-Neelx-2}
\end{figure}

The expressions of the correlation functions for
four-spin plaquette are given by
\be \left\{\begin{array}{c}
 {\mathcal C}_{\textrm{N\'eel}}^z (n)=0,\\[0.9em]
    {\mathcal C}_{\textrm{N\'eel}}^x (n)=\frac{(\zeta_1+1)^2}{8(\zeta_1^2+1)},\\[0.9em]
    {\mathcal C}_{\textrm{N\'eel}}^y (n)=\frac{(\zeta_1-1)^2}{8(\zeta_1^2+1)},
\end{array}\right.
\label{CNx-1-plaquette}\ee
with $\zeta_1=\frac{\gamma J}{h_{\rm z}-J+J'/2-E_1}$.
These relations hold for $n \ne 0$, since
${\mathcal C}_{\textrm{N\'eel}}^\alpha(n=0)=1/4$. Otherwise,
they are independent of $n$, which is true only for $N=4$. 
Because of the PBC, correlation functions satisfy the relation,
${\mathcal C}_{\textrm{N\'eel}}^\alpha (N-n)={\mathcal C}_{\textrm{N\'eel}}^\alpha (n)$.
The result shows that ${\mathcal C}_{\textrm{N\'eel}}^x (n)
={\mathcal C}_{\textrm{N\'eel}}^y (n)$, when $\gamma=0$.
Variation of ground state energy per site,
$E_{\rm G}$, and ${\mathcal C}_{\textrm{N\'eel}}^x (n=1)$
with respect to $J'/J$ for this
plaquette are shown by dashed line
in Fig. \ref{ground-state-energy-correlation}, 
when $\gamma=1$ and $h_{\rm z}=0$.

The results of four-spin plaquette has no importance
in general since they are
drastically different from their respective values
for the system of thermodynamic limit, $N\rightarrow \infty$,
specifically for the region where the strong quantum fluctuation persists.
But symmetries of the physical quantities for the
spin plaquette as obtained here remain unaltered 
in the thermodynamic limit. 
Therefore, they provide useful clue for obtaining
results of larger system. For example, 
Eq. \ref{CNx-1-plaquette} shows that
${\mathcal C}_{\textrm{N\'eel}}^z (n)=0$, for any values of the
parameters. It is found true in the asymptotic limit. 
Further, ${\mathcal C}_{\textrm{N\'eel}}^x (n)={\mathcal C}_{\textrm{N\'eel}}^y (n)$,
when $\gamma=0$, and moreover, 
 ${\mathcal C}_{\textrm{N\'eel}}^x (n,\pm \gamma)
={\mathcal C}_{\textrm{N\'eel}}^y (n,\mp \gamma)$, for any values of $J'$
and $h_{\rm z}$. It corresponds to the fact that 
values of ${\mathcal C}_{\textrm{N\'eel}}^\alpha (n),\,\alpha=x,y$, 
are interchangeable about  the point, $\gamma=0$. 
Subsequent studies reveal that these properties are 
independent of $N$ by virtue of the symmetry. 
Finally it leads to the fact that
no LRO is there when $\gamma=0$, irrespective of the values of
$J'$ and $h_{\rm z}$, as pointed out before in cases {\it 3}
and {\it 5}. 

In this context, $E_{\rm G}$, and ${\mathcal C}_{\textrm{N\'eel}}^x (n=1)$ are
compared with the exact analytic and numerical results as displayed 
in Fig. \ref{ground-state-energy-correlation} (a) and (b), respectively.
It reveals that value of both $E_{\rm G}$, and ${\mathcal C}_{\textrm{N\'eel}}^x (1)$
for the plaquette is equal to the respective exact results when
$\gamma=1$, but only for $h_{\rm z}=0$, and $J'=0$. It happens due to
the fact at this point (case {\it 1}), no quantum fluctuation persists.
As a result, both
the values of $E_{\rm G}$, and ${\mathcal C}_{\textrm{N\'eel}}^x (1)$ 
attain their respective maximum value. So,
${\mathcal C}_{\textrm{N\'eel}}^x (1)$ touches its saturated value,
{\it i. e.}, ${\mathcal C}_{\textrm{N\'eel}}^x (1)=1/4$, at this point.
It is another instance where important results of the
larger system could have been
captured in its four-spin replica. 
\subsection{Numerical results}
In order to detect the existence of AFM LRO
in the ground state of $H$,  
$x$-component of staggered spin correlations,
${\mathcal C}_{\textrm{N\'eel}}^x (n)$ has been computed.
No LRO other than
${\mathcal C}_{\textrm{N\'eel}}^x (n)$, (like FM and chiral phases) 
are found to appear here. 
In order to investigate the properties of
${\mathcal C}_{\textrm{N\'eel}}^x (n)$,  
ground states of spin chains with sites
$N=20,24,$ and 28 are obtained using the Lanczos exact diagonalization
techniques. As the Hamiltonian does not preserve the $U(1)$
symmetry, Hilbert space accommodates the states of all 
$S_{\rm T}^z$ sectors. So the Hamiltonian matrix has been 
spanned in the extended Hilbert space comprising with
$S_{\rm T}^z=\pm N/2, \pm (N-1)/2,\cdots, 0$, sectors.
Ultimately Hilbert space is reduced manifold by
taking into account the translational symmetry of one lattice
unit. Therefore momentum wave vector $k$ is introduced to
associate this symmetry by invoking PBC.
The unique ground state corresponds to either
$k=0$ or $k=\pi$, depending on the values of $N$,
and other parameters.

The variation of ${\mathcal C}_{\textrm{N\'eel}}^x (n)$ 
with respect to $n$, $J'/J$, and $h_{\rm z}$
has been studied extensively, which is 
shown in Fig \ref{Correlation-Neelx-1} and
\ref{Correlation-Neelx-2}, for $\gamma =1$ and 2,
respectively, when $N=28$. For each value of
$\gamma$, ${\mathcal C}_{\textrm{N\'eel}}^x (n)$ is
plotted for $h_{\rm z}=-2,\,-1,\,0,\,1,\,2$. In
Figs \ref{Correlation-Neelx-1} (c) and
\ref{Correlation-Neelx-2} (c), ${\mathcal C}_{\textrm{N\'eel}}^x (n)$
are shown for $\gamma =1$ and 2, when $h_{\rm z}=0$.
They indicate that AFM LRO exists in the region, 
$-2<J'/J<2$, extended uniformly around its center 
$J'/J=0$. It confirms the fact that QPT occurs in this model
at the points $J'/J=-2,\,+2$, even in the absence of
magnetic field. 
The value of ${\mathcal C}_{\textrm{N\'eel}}^x (n)$
is also found symmetric about the point, $J'/J=0$. 
In the regime, $-2<J'/J<2$, correlation function 
is very close to its saturated value,
${\mathcal C}_{\textrm{N\'eel}}^x (n)\approx 1/4$,
for both $\gamma =1$ and 2. With the increase (decrease) of
$h_{\rm z}$, the region hosting LRO shifts toward more positive
(negative) side of $J'/J$ without changing
the width of the region. The diagrams
\ref{Correlation-Neelx-1} (d) and
\ref{Correlation-Neelx-2} (d), indicate that
LRO exists in the regime, $0<J'/J<4$, when $h_{\rm z}=1$.
Similarly,  diagrams
\ref{Correlation-Neelx-1} (e) and
\ref{Correlation-Neelx-2} (e), show that
LRO persists in the regime, $2<J'/J<6$, when $h_{\rm z}=2$.

On the other hand, AFM LRO is found to exist in the regions,
$-4<J'/J<0$, and $-6<J'/J<-2$, when $h_{\rm z}=-1$, and
$h_{\rm z}=-2$, respectively, as depicted in two pairs of diagrams,
\ref{Correlation-Neelx-1} (b), 
\ref{Correlation-Neelx-2} (b), and
\ref{Correlation-Neelx-1} (a), 
\ref{Correlation-Neelx-2} (a). 
 The profile of ${\mathcal C}_{\textrm{N\'eel}}^x (n)$  is absolutely
symmetric around the center of the region ($J'/J=0$)
when $h_{\rm z}=0$, whereas for $h_{\rm z} \ne 0$, 
those shapes become more asymmetric around the respective 
centres of those region for nonzero LRO. 
Another interesting feature of these diagrams is that
value of ${\mathcal C}_{\textrm{N\'eel}}^x (n)$ decreases
with the increase of $|h_{\rm z}|$. This is due to the fact
that contribution of Zeeman energy to the ground state
increases with the increase of $|h_{\rm z}|$ at the
expense of exchange energy. Since the origin of 
magnetic LRO is associated with the cooperative exchange energy,
increase of $|h_{\rm z}|$ leads to the lowering
of the values of ${\mathcal C}_{\textrm{N\'eel}}^x (n)$,
as the contribution of exchange energy is becoming less.
However, magnetic LRO is not at all diminished by the magnetic field
as long as $h_{\rm z}/J'$ is finite. But the region of LRO is found to 
shift as an effect of the field.  
Hence the region of nonzero magnetic LRO can be identified by the
relation, $2(h_{\rm z}-J)<J'<2(J+h_{\rm z})$,
for arbitrary $h_{\rm z}$.
${\mathcal C}_{\textrm{N\'eel}}^x (n)$ is found to diminish exponentially with
the separation $n$ outside this region which confirms the absence of LRO. 
Furthermore, in order to mark the boundary
of this region sharply, Binder cumulant for the staggered
correlation function has been evaluated numerically as described below.

\begin{figure}[h]
  \psfrag{BNx}{ ${\mathcal B}_{\textrm{N\'eel}}^x$}
    \psfrag{a}{(a)}
    \psfrag{b}{(b)}
  \psfrag{c}{(c)}
  \psfrag{d}{(d)}
  \psfrag{e}{(e)}
  \psfrag{d1}{$\gamma\!=\!1$}
  \psfrag{jp}{ $J'/J$}
   \psfrag{x}{ $2/3$}
  \psfrag{N=28}{$N\!\!=\!\!28$}
\psfrag{N=20}{$N\!\!=\!\!20$}
\psfrag{N=24}{$N\!\!=\!\!24$}
\psfrag{N=12}{$N\!\!=\!\!12$}
\psfrag{N=16}{$N\!\!=\!\!16$}
\psfrag{h=0}{$h_{\rm z}\!=\!0$}
\psfrag{h=1}{$h_{\rm z}\!=\!1$}
\psfrag{h=2}{$h_{\rm z}\!=\!2$}
\psfrag{h=-1}{$h_{\rm z}\!=\!-1$}
\psfrag{h=-2}{$h_{\rm z}\!=\!-2$}
\psfrag{0}{0}
\psfrag{1}{1}
\psfrag{2}{2}
\psfrag{-1}{$-1$}
\psfrag{-3}{$-3$}
\psfrag{-2}{$-2$}
\psfrag{-4}{$-4$}
\psfrag{-5}{$-5$}
\psfrag{-6}{$-6$}
\psfrag{-7}{$-7$}
\psfrag{3}{3}
\psfrag{4}{4}
\psfrag{5}{5}
\psfrag{6}{6}
\psfrag{7}{7}
\psfrag{0.1}{$0.1$}
\psfrag{0.2}{$0.2$}
\psfrag{0.3}{$0.3$}
\psfrag{0.4}{$0.4$}
\psfrag{0.0}{$0.0$}
\psfrag{0.5}{$0.5$}
\psfrag{0.6}{$0.6$}
\psfrag{0.7}{$0.7$}
\includegraphics[width=230pt]{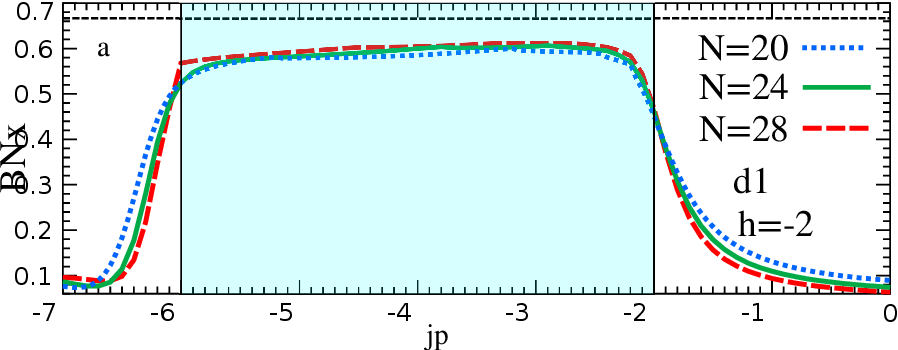}
\includegraphics[width=230pt]{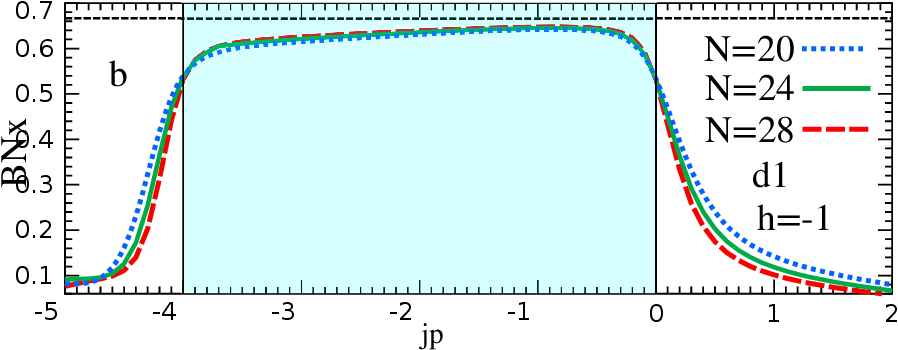}
\includegraphics[width=230pt]{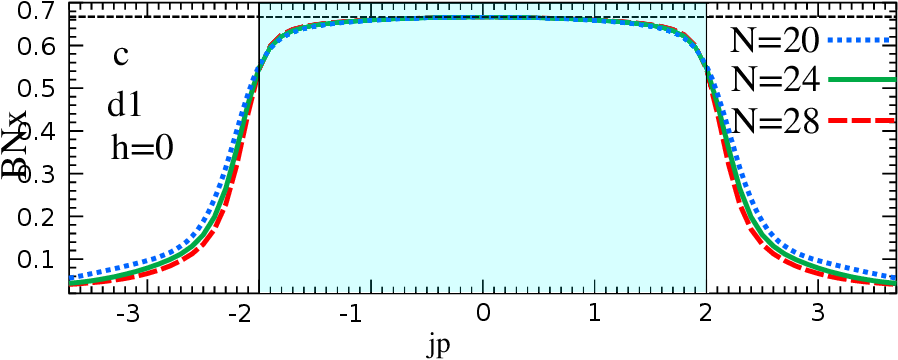}
\includegraphics[width=230pt]{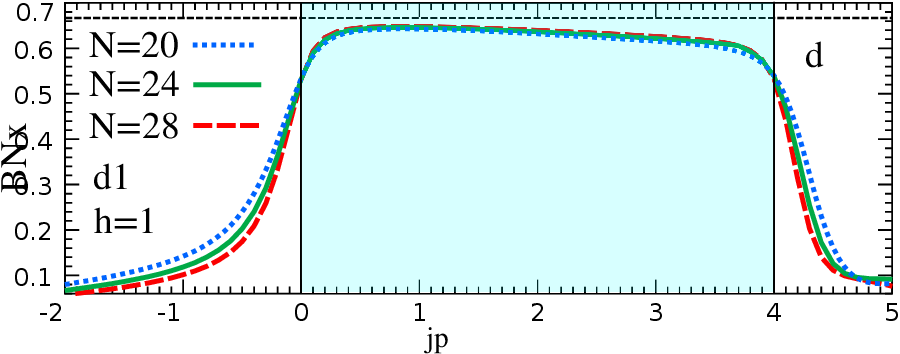}
\includegraphics[width=230pt]{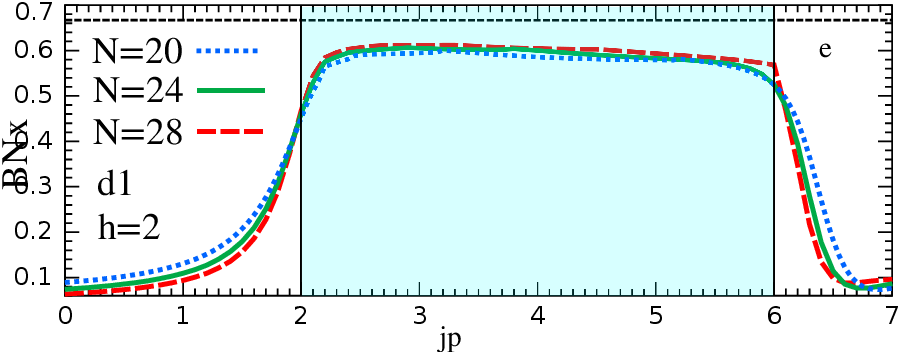}
\caption{Binder cumulant when $\gamma\!=\!1$ for  $h_{\rm z}=-2$ (a),
  $h_{\rm z}=-1$ (b), $h_{\rm z}=0$ (c), $h_{\rm z}=1$ (d), and $h_{\rm z}=2$ (e).}
 \label{Binder-cumulant-1}
\end{figure}

\begin{figure}[h]
  \psfrag{BNx}{ ${\mathcal B}_{\textrm{N\'eel}}^x$}
    \psfrag{a}{(a)}
    \psfrag{b}{(b)}
  \psfrag{c}{(c)}
  \psfrag{d}{(d)}
  \psfrag{e}{(e)}
  \psfrag{d2}{$\gamma\!=\!2$}
  \psfrag{jp}{ $J'/J$}
   \psfrag{x}{ $2/3$}
  \psfrag{N=28}{$N\!\!=\!\!28$}
\psfrag{N=20}{$N\!\!=\!\!20$}
\psfrag{N=24}{$N\!\!=\!\!24$}
\psfrag{N=12}{$N\!\!=\!\!12$}
\psfrag{N=16}{$N\!\!=\!\!16$}
\psfrag{h=0}{$h_{\rm z}\!=\!0$}
\psfrag{h=1}{$h_{\rm z}\!=\!1$}
\psfrag{h=2}{$h_{\rm z}\!=\!2$}
\psfrag{h=-1}{$h_{\rm z}\!=\!-1$}
\psfrag{h=-2}{$h_{\rm z}\!=\!-2$}
\psfrag{0}{0}
\psfrag{1}{1}
\psfrag{2}{2}
\psfrag{-1}{$-1$}
\psfrag{-3}{$-3$}
\psfrag{-2}{$-2$}
\psfrag{-4}{$-4$}
\psfrag{-5}{$-5$}
\psfrag{-6}{$-6$}
\psfrag{-7}{$-7$}
\psfrag{3}{3}
\psfrag{4}{4}
\psfrag{5}{5}
\psfrag{6}{6}
\psfrag{7}{7}
\psfrag{0.1}{$0.1$}
\psfrag{0.2}{$0.2$}
\psfrag{0.3}{$0.3$}
\psfrag{0.4}{$0.4$}
\psfrag{0.0}{$0.0$}
\psfrag{0.5}{$0.5$}
\psfrag{0.6}{$0.6$}
\psfrag{0.7}{$0.7$}
\includegraphics[width=230pt]{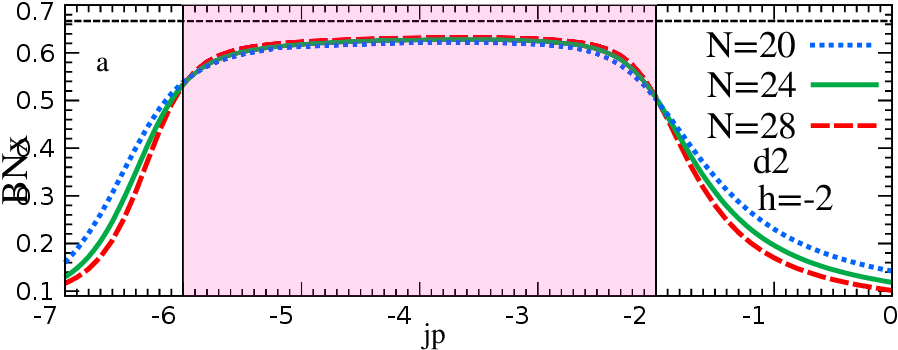}
\includegraphics[width=230pt]{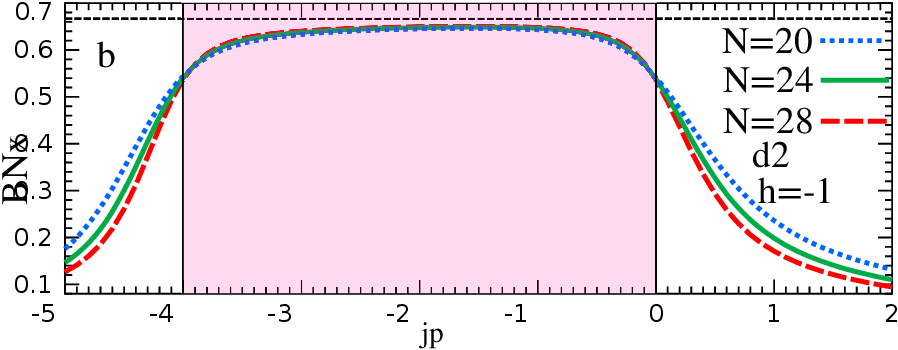}
\includegraphics[width=230pt]{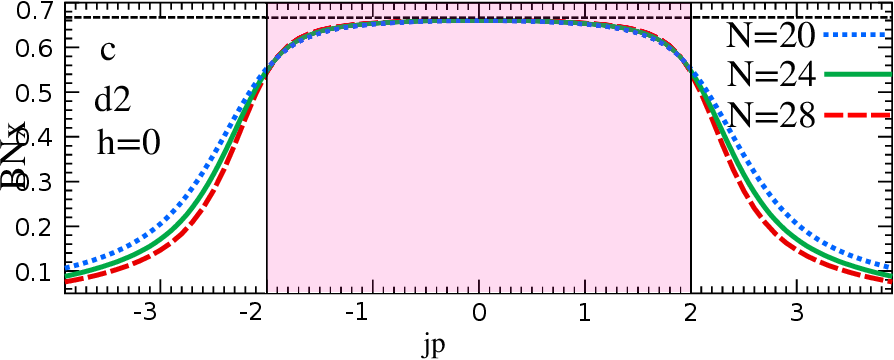}
\includegraphics[width=230pt]{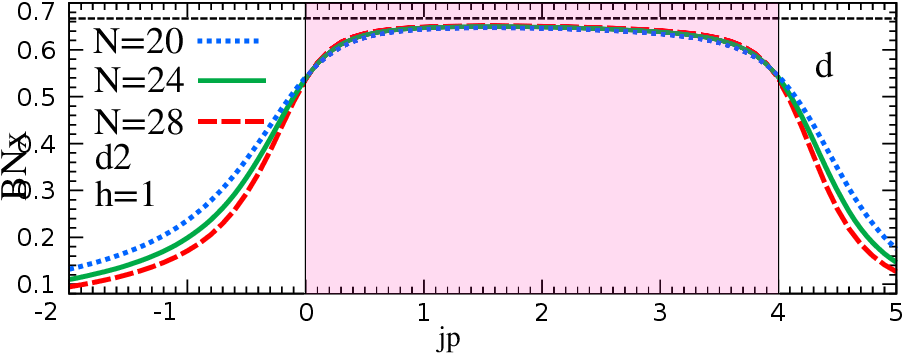}
\includegraphics[width=230pt]{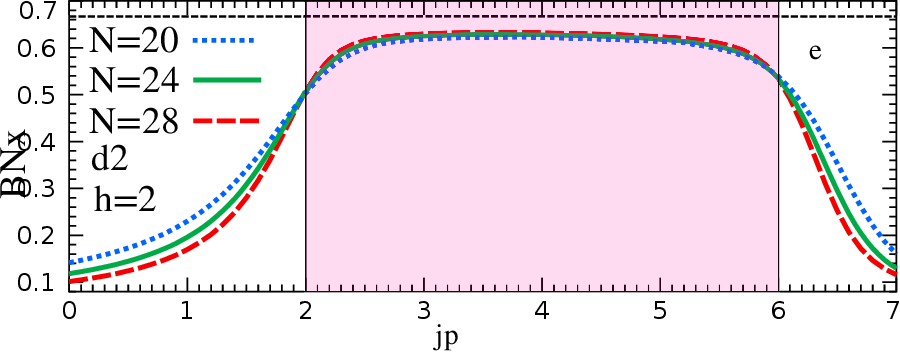}
\caption{Binder cumulant, ${\mathcal B}_{\textrm{N\'eel}}^x$,
  when $\gamma=2$, for  $h_{\rm z}=-2$ (a), $h_{\rm z}=-1$ (b),
  $h_{\rm z}=0$ (c), $h_{\rm z}=1$ (d), and $h_{\rm z}=2$ (e).}
 \label{Binder-cumulant-2}
\end{figure}

Evolution of Binder cumulant, ${\mathcal B}_{\textrm{N\'eel}}^x(N)$, (Eq. \ref{BP})
with respect to the number of sites, $N$ helps to
identify the region where LRO exists. 
For this purpose, value of  ${\mathcal B}_{\textrm{N\'eel}}^x(N)$,
for a system of finite number of spins, $N=20,\,24,\,28$, have been
estimated for different sets of parameters, 
$h_{\rm z}=-2,\,-1,\,0,\,1,\,2$, and $\gamma =1$ and 2. 
If there exists LRO, ${\mathcal B}_{\textrm{N\'eel}}^x(N)$
grows with the increase of system size, $N$, in contrast, 
${\mathcal B}_{\textrm{N\'eel}}^x(N)$ decays with the increase $N$, 
where short-ranged order persists\cite{Binder1,Binder2}. 
As a result, at the transition points, 
value of ${\mathcal B}_{\textrm{N\'eel}}^x(N)$ remains unaltered 
for any values of $N$, which are
shown in Figs \ref{Binder-cumulant-1} for $\gamma =1$,
and \ref{Binder-cumulant-2} for  $\gamma =2$, 
in great details.  

According to the definition of
${\mathcal B}_{\textrm{N\'eel}}^x$, (Eq. \ref{BP}),
the maximum possible value of ${\mathcal B}_{\textrm{N\'eel}}^x$ is
2/3, which is shown by the horizontal dashed line
in the Figs \ref{Binder-cumulant-1} and \ref{Binder-cumulant-2}.
It will be observed when the corresponding magnetic order will
attain its saturated value. In this model, it happens 
for $h_{\rm z}=0$, and $J'/J=0$,
as shown in Figs \ref{Binder-cumulant-1} and \ref{Binder-cumulant-2} (c), where
the peak of ${\mathcal B}_{\textrm{N\'eel}}^x$ touches the horizontal dashed line 
(maximum value) for any value of $N$. It occurs due to the fact that
at this point, $\langle M_{\textrm{N\'eel}}^x\rangle=1/2$, which is equal to its
saturated value for spin-1/2 systems and corresponds to the
maximum value of correlation function,
${\mathcal C}_{\textrm{N\'eel}}^x (n)=1/4$, for arbitrary $n$,
as they are related by the Eq. \ref{MtoC}.
Those results have been derived analytically before for $\gamma=1$
in Eq. \ref{eigenvalues-x}. 
As a result, $\langle (M_{\textrm{N\'eel}}^x)^4\rangle=1/16$,
which finally leads to, ${\mathcal B}_{\textrm{N\'eel}}^x=2/3$. 

In contrast, for
$h_{\rm z}\ne 0$, value of ${\mathcal B}_{\textrm{N\'eel}}^x$
tends to increase with the increase of $N$, indicating to touch its
maximum value for higher values of $N$, beyond $N=28$. This phenomenon indicates the
existence of LRO in a specific region. Obviously, the reverse phenomenon
confirms the absence of LRO. 
However, value of the peak for ${\mathcal B}_{\textrm{N\'eel}}^x$
decreases with the increase of $|h_{\rm z}|$, as clearly
depicted in the Figs. \ref{Binder-cumulant-1} and \ref{Binder-cumulant-2},
(a, b, d, e). This phenomenon can be explained 
by examining the relations among the relevant quantities, ${\mathcal B}_{\textrm{N\'eel}}^x$,
$M_{\textrm{N\'eel}}^x$,  and  ${\mathcal C}_{\textrm{N\'eel}}^x (n)$ as stated
in the Eqs \ref{BP}, \ref{Mag} and \ref{MtoC}, respectively.
With the increase of $|h_{\rm z}|$, contribution of Zeeman energy
to the ground state of the system increases. The competing Zeeman term 
reduces the effect of exchange term, which in turn reduces the value of 
$\langle M_{\textrm{N\'eel}}^x\rangle$ below to its saturated value (1/2) in such a
fashion that ${\mathcal B}_{\textrm{N\'eel}}^x<2/3$. 

\begin{figure}[h]
    \psfrag{a}{(a)}
    \psfrag{b}{(b)}
  \psfrag{BNx}{ ${\mathcal B}_{\textrm{N\'eel}}^x$}
  \psfrag{BNz}{ ${\mathcal B}_{\textrm{N\'eel}}^z$}
  \psfrag{BCz}{${\mathcal B}_{\rm Chiral}^z$}  
  \psfrag{d1}{$\gamma\!=\!1$}
\psfrag{jp}{ $J'/J$}
\psfrag{h}{$h_{\rm z}$}
\psfrag{0}{$0$}
\psfrag{2}{$2$}
\psfrag{1.0}{$1.0$}
\psfrag{2.0}{$2.0$}
\psfrag{-2.0}{$-2.0$}
\psfrag{-0.1}{$-0.1$}
\psfrag{-1.0}{$-1.0$}
\psfrag{-1.5}{$-1.5$}
\psfrag{-0.5}{$-0.5$}
\psfrag{1.5}{$1.5$}
\psfrag{-1}{$-1$}
\psfrag{-3}{$-3$}
\psfrag{-2}{$-2$}
\psfrag{-4}{$-4$}
\psfrag{-5}{$-5$}
\psfrag{-6}{$-6$}
\psfrag{ 2}{$2$}
\psfrag{ 4}{$4$}
\psfrag{ 0}{$0$}
\psfrag{ 6}{$6$}
\psfrag{-7}{$-7$}
\psfrag{-8}{$-8$}
\psfrag{3}{3}
\psfrag{4}{$4$}
\psfrag{5}{5}
\psfrag{6}{$6$}
\psfrag{8}{$8$}
\psfrag{7}{7}
\psfrag{0.01}{$0.01$}
\psfrag{0.02}{$0.02$}
\psfrag{0.03}{$0.03$}
\psfrag{-0.01}{$-0.01$}
\psfrag{-0.02}{$-0.02$}
\psfrag{-0.03}{$-0.03$}
\psfrag{0.00}{$0.00$}
\psfrag{-0.04}{$-0.04$}
\psfrag{-0.05}{$-0.05$}
\psfrag{0.1}{$0.1$}
\psfrag{0.2}{$0.2$}
\psfrag{0.3}{$0.3$}
\psfrag{0.4}{$0.4$}
\psfrag{0.0}{$0.0$}
\psfrag{0.5}{$0.5$}
\psfrag{0.6}{$0.6$}
\psfrag{0.7}{$0.7$}
\psfrag{0.01}{$0.01$}
\psfrag{0.02}{$0.02$}
\psfrag{0.03}{$0.03$}
\includegraphics[width=230pt,angle=0]{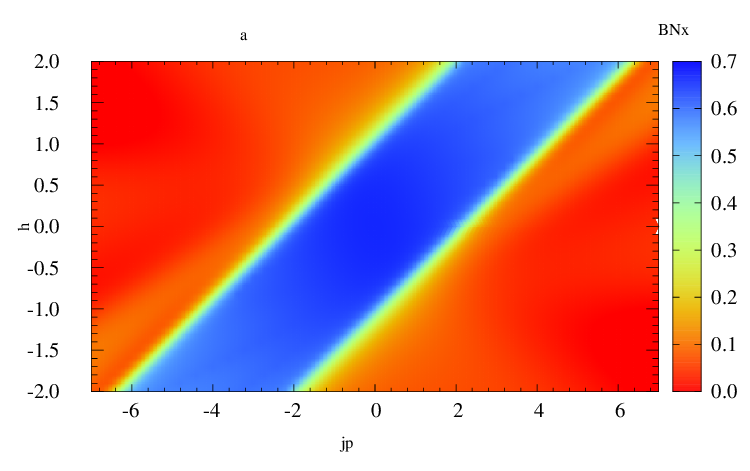}
\vskip -0.50 cm
\includegraphics[width=145pt,angle=-90]{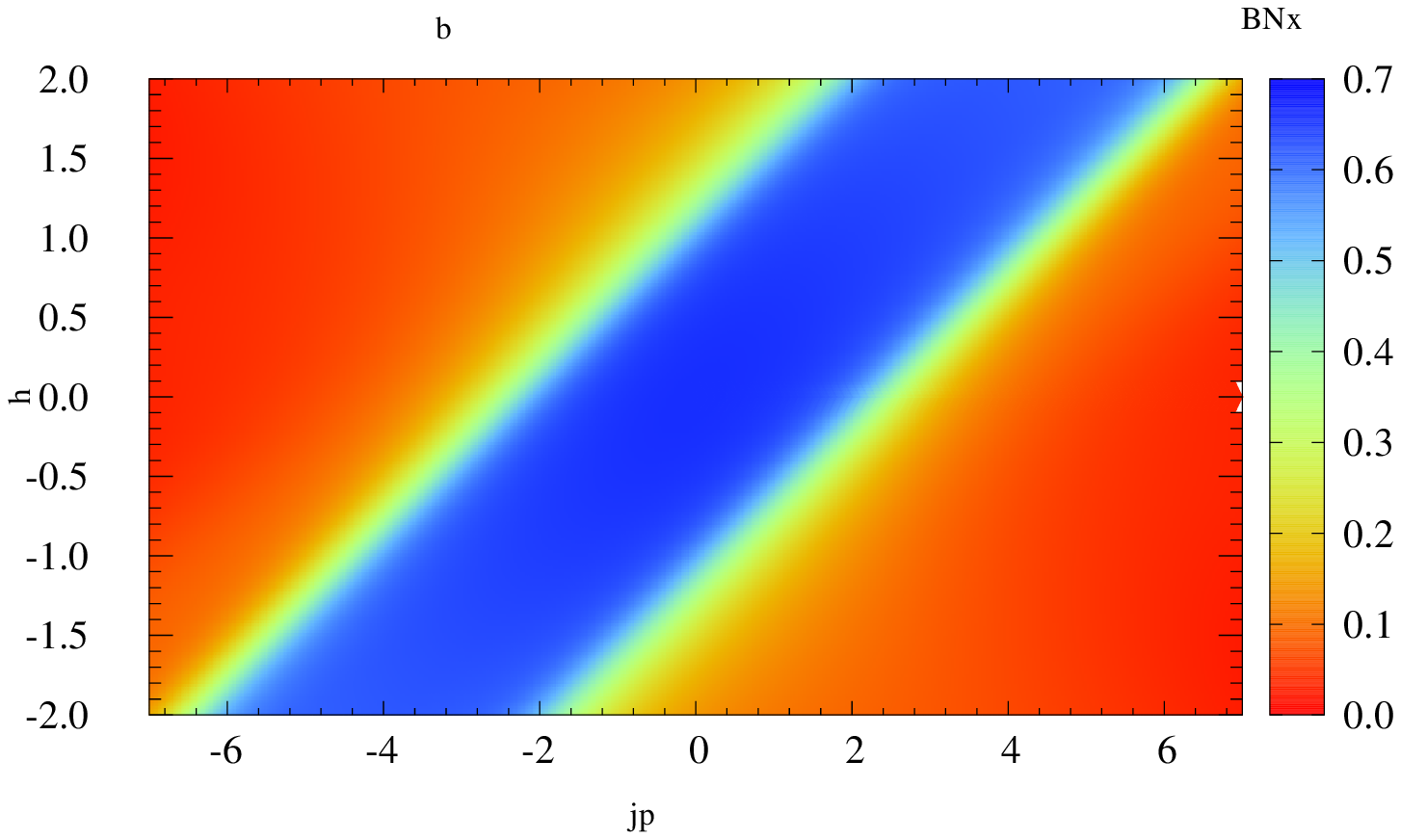}
\caption{Density plot of Binder cumulant, ${\mathcal B}_{\textrm{N\'eel}}^x$,
  when $\gamma=1$ (a), and
     $\gamma=2$ (b), for $-6\le J'/J\le 6$.}
 \label{density-plot}
\end{figure}

${\mathcal B}_{\textrm{N\'eel}}^x(N)$ for $N=20,\,24$ and 28 are drawn with
red dashed, green solid and blue dotted lines in each diagrams of Figs.
\ref{Binder-cumulant-1} and \ref{Binder-cumulant-2}. 
The shaded region bounded by two
vertical lines indicates the regime where LRO exists. 
The region for nonzero LRO corresponds to 
the relation, $2(h_{\rm z}-J)<J'<2(J+h_{\rm z})$, for both $\gamma =1$ and 2.
Density plot for ${\mathcal B}_{\textrm{N\'eel}}^x$ when
$\gamma =1$ and 2 are shown in Fig. \ref{density-plot} (a) and (b),
respectively. 
It indicates LRO is not affected by the value of $\gamma$ as long as it is
nonzero and finite. Analytic derivation of the correlation functions,
${\mathcal C}_{\textrm{N\'eel}}^x (n)$, in terms of JW fermionization
have been carried out in the next section. 
The numerical results obtained in this section
show an excellent agreement with the analytic counterpart.


\subsection{Jordan-Wigner fermionization}
Several properties of the Hamiltonian, {\it e. g.},
dispersion relations, ground state energy, energy gap and spin-spin 
correlations for $H$ have been obtained
analytically in this section after expressing
$H$ in terms of spinless fermions. 
Under JW transformations\cite{JW}:
\be\left\{
\begin{aligned}
S_j^+&=c_j^\dag\prod_{l=1}^{j-1}(1-2\,n_l),\\
S_j^-&=\prod_{l=1}^{j-1}(1-2\,n_l)\,c_j,\\
S_j^z&=n_j-\frac{1}{2},
 \label{JW}
\end{aligned}\right.
\ee
where $n_l=c_l^\dag c_l$, is the number operator, 
the Hamiltonian 
looks like
\bea
   H&=&\!\frac{1}{2}\sum_j\!\bigg[J\left(c_j^\dag c_{j+1}\!+\!c_{j+1}^\dag c_j\right)
     -\frac{J'}{2}\left(c_j^\dag c_{j+2}\!+\!c_{j+2}^\dag c_j\right)\nonumber\\[-0.2em]
     && \;\;  +J\gamma\left( c_j^\dag c_{j+1}^\dag+c_{j+1} c_j\right)
     +2h_{\rm z}\left(c_j^\dag c_{j}-\frac{1}{2}\right)\bigg].
 \label{hamJW}
 \eea
Comparing this to the form of Kitaev 1D model\cite{Kitaev},
it reveals that $J$ and $J'$ turn out to be
equivalent to the NN and NNN hopping integrals, 
respectively, while $\gamma$ 
and $h_{\rm z}$ serve as the superconducting and 
chemical potentials, respectively. 
%
Under the Fourier transformation, 
\[c_j=\frac{1}{\sqrt N}\sum_{k\in{\rm BZ}}c_{k}\,e^{-i{k}j},\]
 Hamiltonian acquires the
Bogoliubov-de Gennes (BdG) form, so, 
\be
H=\sum_{k}\left[\psi_{k}^\dag\, \mathcal H({k})\,\psi_{k}+\epsilon_{k}-\frac{h_{\rm z}}{2}\right],
\ee
where $\psi_{k}^\dag=\left[c^\dag_{k}\;\,c_{-k}\right]$,
$\epsilon_{k}=\frac{J}{2}\!\left(\cos{(k)} -
\frac{J'}{2J}\cos{(2k)}\right)+h_{\rm z}/2$. 
Now introducing the vectors,
$\boldsymbol g=(g_x,g_y,g_z)$, and
$\boldsymbol \sigma=(\sigma_x,\sigma_y,\sigma_z)$,
where $\sigma_\alpha,\;\alpha=x,y,z$, are the Pauli matrices, 
one can write 
$\mathcal H({k})=\boldsymbol g \cdot \boldsymbol \sigma$.  
Now, 
\[\left\{\begin{array}{l}
g_x=0,\\ [0.2em]
g_y=\Delta_{k},\\ [0.2em]
g_z=\epsilon_{k},\end{array}\right. \]
where $\Delta_{k}=\frac{1}{2}J\gamma\sin{(k)}$.

Under the Bogoliubov transformation,
$\phi_{k}=\mathcal P_{k}\psi_{k}$,
where $\mathcal P_{k}=\left(\begin{array}{cc}u_{k}&-v_{k}\\
  v^\star_{k}&u^\star_{k}\end{array}\right)$, and 
$\phi_{k}^\dag=\left[\gamma^\dag_{k}\;\gamma_{-k}\right]$, 
the Hamiltonian becomes 
\bea
H&=&\sum_{k}\left[\phi_{k}^\dag\,\left(\mathcal P_{k}^{-1}\right)^\dag \mathcal H({k})\,\mathcal P_{k}^{-1}\phi_{k}+\epsilon_{k}\right],\nonumber\\
&=&NE_{\rm G}+\sum_{k}E_{k}\left[\gamma_{k}^\dag\,\gamma_{k}+\gamma_{-k}^\dag\,\gamma_{-k}\right],\nonumber
\eea
when $|u_{k}|=\sqrt{\left(1+\epsilon_{k}/E_{k} \right)/2}$,
$|v_{k}|=\sqrt{\left(1-\epsilon_{k}/E_{k} \right)/2}$, 
where dispersion relation of the Bogoliubon (quasiparticle) excitation is
\be E_{k}=\sqrt{\epsilon_{k}^2+|\Delta_{k}|^2},\ee 
and the ground state energy per site is
\be
E_{\rm G}=-\frac{1}{2\pi}\int_{-\pi}^{+\pi}E_{k} d{k}.
\label{Eq-Eg}
\ee
It indicates that superconducting phase survives as long as $\gamma \ne 0$,
where Cooper pairs are formed between parallel spins. 
Value of ground state energy obtained numerically is compared with
the exact result, $E_{\rm G}$ (Eq. \ref{Eq-Eg}) which is shown in
Fig. \ref{ground-state-energy-correlation} (a), 
when $\gamma=1$, and $-3\le J'/J\le 3$,
for $h_{\rm z}=0$ .
Eq. \ref{Eq-Eg} is plotted in black line
while red circles are the numerical data. 
The excellent agreement confirms the extreme accuracy of the
numerical result. 

However, for $\gamma = 0$, 
dispersion relation of spin excitation
is gapless as the Hamiltonian becomes
\be
H(\gamma = 0)=2\sum_{k}\epsilon_{k}c_{k}^\dag c_{k},\nonumber
\ee
since in this case  $|u_{k}|=1$ and $|v_{k}|=0$.
The ground state energy per site for $h_{\rm z} = 0$, is estimated now by summing the
negative energy states\cite{Japaridze},
\be E_{\rm G}=\frac{1}{\pi}\int_{k\in(\epsilon_{k}<0)}\epsilon_{k}\,d{k}.\ee
Magnetic LRO, along with the topological superconducting phase
cease to exist in this case. 
Dispersion relations ($E_{k}$) and energy gap ($E_{\rm Gap}$)
of the system have been evaluated
for various cases as described below.
$E_{\rm Gap}$ corresponds to the minimum value of
$E_{k}$, which has been obtained by solving the
equation, $\frac{dE_{k}}{dk}=0$, numerically, for the
fixed values of other parameters. 
Value of $E_{\rm Gap}$ is important
while determining the topological phase transition points. Because at
the transition point $E_{\rm Gap}$ must vanish. 
\subsection{Dispersion relations}
\begin{figure}[h]
   \psfrag{d0}{ $\gamma=0$}
      \psfrag{h0}{ $h_{\rm z}=0$} 
\psfrag{p}{\large  $-\pi$}
\psfrag{q}{\large  $-\frac{\pi}{2}$}
\psfrag{n}{\large $+\frac{\pi}{2}$}
\psfrag{m}{\large $+\pi$}
\psfrag{e}{$\epsilon_{k}/J$}
\psfrag{g}{ $J'/J$}
\psfrag{k}{k}
\psfrag{0}{0}
\psfrag{1}{1}
\psfrag{2}{2}
\psfrag{-1}{$-1$}
\psfrag{-2}{$-2$}
\psfrag{-3}{$-3$}
\psfrag{-4}{$-4$}
\psfrag{-5}{$-5$}
\psfrag{3}{3}
\psfrag{4}{4}
\psfrag{5}{5}
\psfrag{6}{6}
\psfrag{-1.0}{$-1.0$}
\psfrag{-0.5}{$-0.5$}
\psfrag{0.0}{$0.0$}
\psfrag{1.0}{$1.0$}
\psfrag{0.5}{$0.5$}
\psfrag{1.5}{$1.5$}
\psfrag{2.0}{$2.0$}
  \includegraphics[width=240pt]{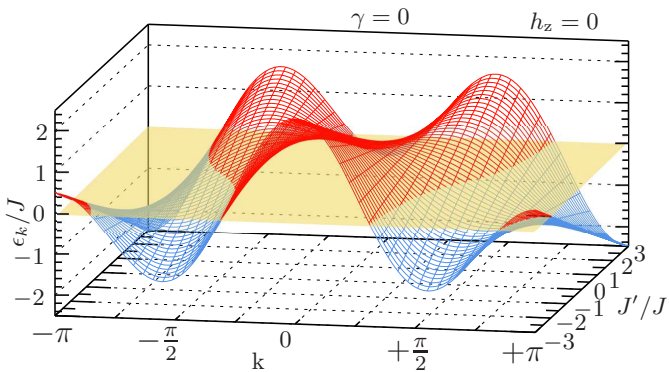}
  \caption{Variation of $\epsilon_{k}/J$ with respect to
    $J'/J$,  within the BZ  for $\gamma=0$, and
    $-3<J'/J<3$, when $h_{\rm z}=0$. Positive and negative portions
    are drawn in red and blue, respectively.}
 \label{dispersion0}
\end{figure}

For $\gamma=0$, and $h_{\rm z}=0$, energy dispersion is always gapless as shown
in Fig. \ref{dispersion0}.  The positive and negative portions of
$\epsilon_{k}$ are
drawn in red and blue, respectively. It shows that, $\epsilon_{k}$ 
vanishes exactly at two distinct points, 
${k}=\pm \arccos{(J- \sqrt{J^2+2J'^2}/2J')}$, as long as
$J'/J<2$. However for $J'/J>2$, $\epsilon_{k}$ vanishes at two
additional points marked by
${k}=\pm \arccos{(J+\sqrt{J^2+2J'^2}/2J')}$\cite{Japaridze}.
In the presence of magnetic field, energy dispersion shifts along the
vertical direction depending on the sign of $h_{\rm z}$. So the 
extent of positive and negative portions for
$\epsilon_{k}$ shift accordingly when $h_{\rm z}\ne 0$. 

\begin{figure}[th]
  \psfrag{a}{\large  (a)}
  \psfrag{b}{\large (b)}
   \psfrag{c}{\large (c)}
   \psfrag{d0}{ $\gamma\!=\!0$}
   \psfrag{d1}{ $\gamma\!=\!1$}
      \psfrag{h0}{ $h_{\rm z}\!=\!0$}
      \psfrag{h1}{ $h_{\rm z}\!=\!1$}
  \psfrag{h-1}{ $h_{\rm z}\!=\!-1$}     
\psfrag{p}{\large  $-\pi$}
\psfrag{q}{\large  $-\frac{\pi}{2}$}
\psfrag{n}{\large $+\frac{\pi}{2}$}
\psfrag{m}{\large $+\pi$}
\psfrag{e}{ $E_{k}/J$}
\psfrag{g}{$J'/J$}
\psfrag{k}{\large k}
\psfrag{0}{0}
\psfrag{1}{1}
\psfrag{2}{2}
\psfrag{-1}{$-1$}
\psfrag{-2}{$-2$}
\psfrag{-3}{$-3$}
\psfrag{-4}{$-4$}
\psfrag{-5}{$-5$}
\psfrag{3}{3}
\psfrag{4}{4}
\psfrag{5}{5}
\psfrag{6}{6}
\psfrag{-1.0}{$-1.0$}
\psfrag{-0.5}{$-0.5$}
\psfrag{0.0}{$0.0$}
\psfrag{1.0}{$1.0$}
\psfrag{0.5}{$0.5$}
\psfrag{1.5}{$1.5$}
\psfrag{2.0}{$2.0$}
  \includegraphics[width=230pt]{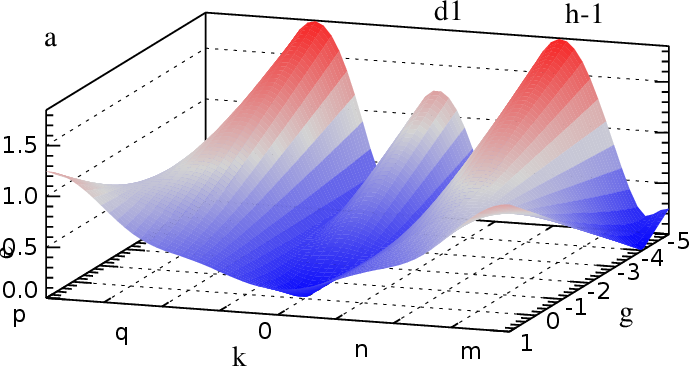}
     \vskip 0.4cm
   \includegraphics[width=230pt]{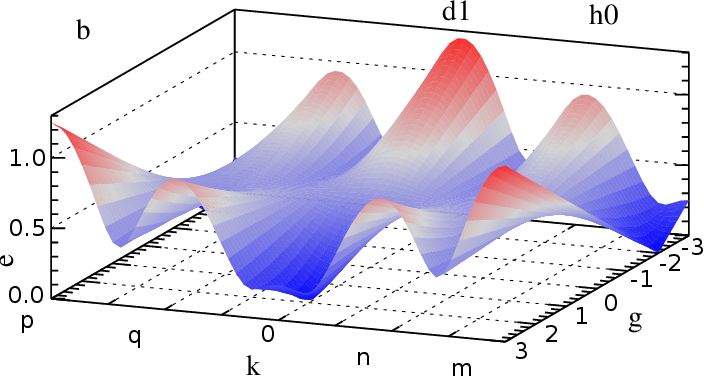}
     \vskip 0.4cm
   \includegraphics[width=230pt]{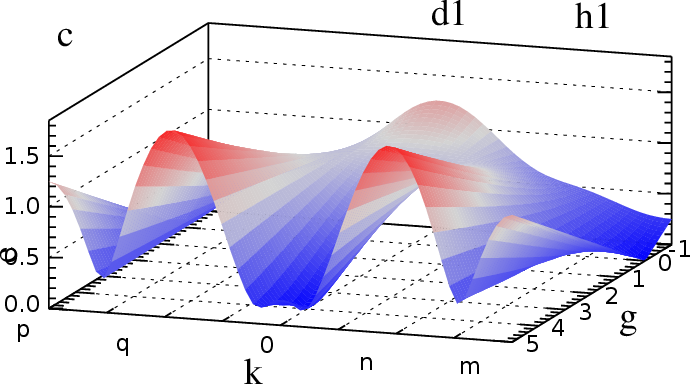}
   \caption{Dispersion relations for $\gamma=1$,
when $h_{\rm z}=-1$ (a), $h_{\rm z}=0$ (b), and $h_{\rm z}=1$ (b).}
 \label{dispersion1}
\end{figure}

In Fig \ref{dispersion1} (a), (b) and (c),
dispersion relations for $h_{\rm z}=-1,\,0$, and $1$ are shown 
respectively, when $\gamma=1$. 
They are qualitatively different. For examples, 
when $h_{\rm z}=0$, three broad peaks and three valleys
are there and they are symmetric about the point $J'/J=0$.
However, the positions of
valleys and peaks are interchanged around $J'/J=0$.
When $J'/J<0$, peaks appear at ${k}=0, \,\pm \pi/2$. 
Height of the
peaks increases with the increase of $|J'/J|$.
Dispersion becomes gapless, $E_{k}=0$, for $J'=-2J$ when k $=\pm \pi$,
and for $J'=2J$ when k = 0. There is always a gap otherwise.

On the other hand, for $h_{\rm z}=1$, there is one large peak
at ${k}=0$, and two small peaks at ${k}=\pm \pi/2$, when
$J'/J<0$. When $J'/J>0$, two large peaks appear
at ${k}=\pm \pi/2$, while the deep valley appear at 
${k}=0$. Gap vanishes when $J'/J=0$ and
${k}=\pm \pi$. Even though the gap function is nonzero since $\gamma\ne 0$,
energy gap vanishes due to the effect of $J'$ and $h_{\rm z}$. 

\begin{figure}[h]
\psfrag{a}{\large  (a)}
\psfrag{b}{\large (b)}
\psfrag{c}{\large (c)}  
\psfrag{gap}{$E_{\rm Gap}/J$}
\psfrag{g}{$E_{\rm Gap}/J$}
   \psfrag{d}{$\gamma$}
   \psfrag{j}{$J'/J$}
\psfrag{h0}{$h_{\rm z}=0$}
\psfrag{h1}{$h_{\rm z}=1$}
\psfrag{h-1}{$h_{\rm z}=-1$}     
\psfrag{0}{0}
\psfrag{1}{1}
\psfrag{2}{2}
\psfrag{-1}{$-1$}
\psfrag{-3}{$-3$}
\psfrag{-2}{$-2$}
\psfrag{-4}{$-4$}
\psfrag{-5}{$-5$}
\psfrag{-6}{$-6$}
\psfrag{3}{3}
\psfrag{4}{4}
\psfrag{5}{5}
\psfrag{6}{6}
\psfrag{0.1}{$0.1$}
\psfrag{0.2}{$0.2$}
\psfrag{0.3}{$0.3$}
\psfrag{0.4}{$0.4$}
\psfrag{0.0}{$0.0$}
\psfrag{0.5}{$0.5$}
\includegraphics[width=230pt]{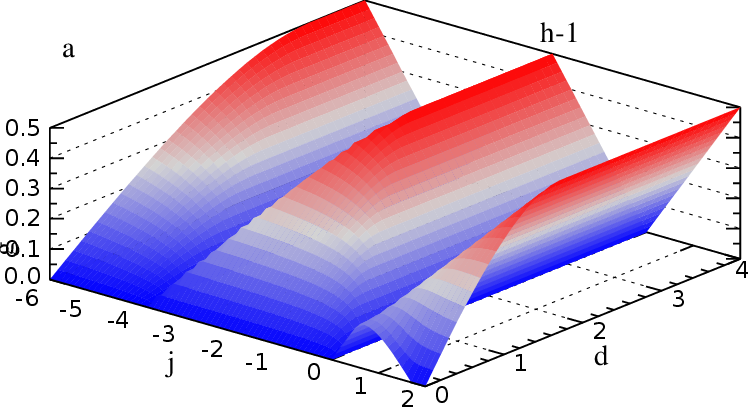}
\vskip 0.1cm
\includegraphics[width=230pt]{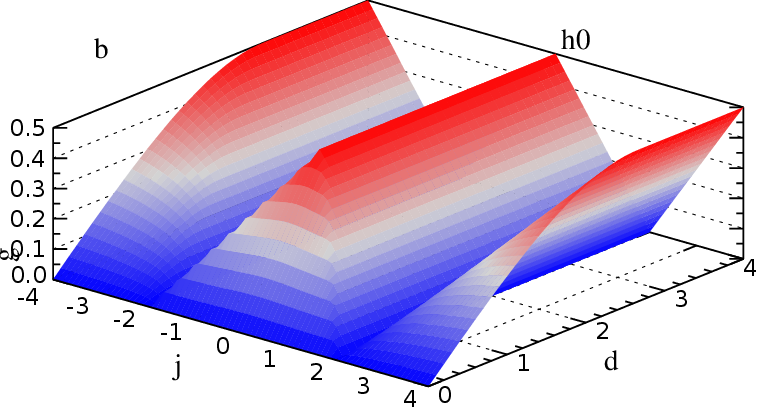}
\vskip 0.1cm
\includegraphics[width=230pt]{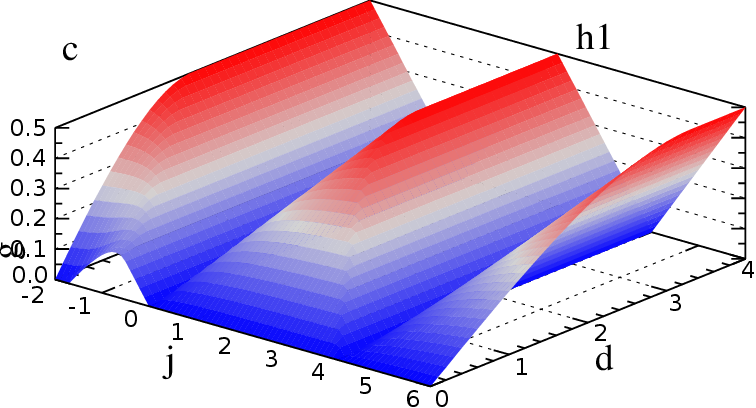}
\caption{Variation of band gap, $E_{\rm Gap}/J$ 
  with respect to both $J'/J$ and $\gamma$, when
$h_{\rm z}=-1$ (a), $h_{\rm z}=0$ (b), and $h_{\rm z}=1$ (c).}
 \label{dispersion-gap}
\end{figure}

In order to identify the gapless region, minimum value of
$E({k})$ has been estimated numerically. True band gap is equal to the
twice of $E_{\rm Gap}$. 
The variation of $E_{\rm Gap}$ in the $\gamma$-$J'/J$ parameter space
is shown in Fig. \ref{dispersion-gap} (a), (b) and (c), 
for $h_{\rm z}=-1,\,0$, and $1$. Obviously, $E_{\rm Gap}=0$,
when $\gamma=0$ and $h_{\rm z}=0$, since the superconducting phase
does not survive as shown in (b). It also serves as a
line over which phase transition occurs.
Additionally $E_{\rm Gap}=0$, along the lines $J'=\pm 2J$, irrespective
of the values of $\gamma$. These lines serve as the boundaries
of the trivial and topological superconducting phases since
the topological phase persists in the annular
region, $-2<J'/J<2$, which will be shown later.

For $\gamma=0$, and $h_{\rm z}\ne 0$, there is a gap in the region,
$-2h_{\rm z}/J<J'/J<0$ when $h_{\rm z}>0$ and in the region
$0<J'/J<2h_{\rm z}/J$, when $h_{\rm z}<0$, as shown in
\ref{dispersion-gap} (c) and (a), respectively, for $h_{\rm z}=1$,
and $h_{\rm z}=-1$.
In contrast, when $\gamma\ne 0$, there is a gap in the spectrum,
except over two parallel lines as described here. For example, 
band gap vanishes at $J'=(-4J,\,0)$, and  $J'=(0,\,4J)$, respectively, 
when $h_{\rm z}=-1$, and $h_{\rm z}=1$ as shown in
(a) and (c), for any value of $\gamma$.
The results indicate that band gap vanishes at
the points, $J'=\pm 2J+2h_{\rm z}$, for arbitrary values of $h_{\rm z}$.
Further, it occurs exactly at ${k}=\pi$, and ${k}=0$ for
$J'=\pm 2J+2h_{\rm z}$, respectively. Otherwise, there is nonzero band gap
and the minimum of band gap appears when ${k}\ne 0,\,\pi$.
Although the value of $E_{\rm Gap}$ depends on $\gamma$, 
location of these boundary lines are insensitive to the value of
$\gamma$. This fact can be understood simply by noting that
 $E_{\rm Gap}$ becomes zero when both $\epsilon_{k}$ and 
$\Delta_{k}$ vanish simultaneously. $\epsilon_{k}$ vanishes
when $k=0$ or $\pi$ irrespective of any value of $\gamma$.
As a result, $\Delta_{k}$ vanishes when $J+h_{\rm z}=\pm J'/2$.

\begin{figure}[h]
  \psfrag{CNx}{\large ${\mathcal C}_{\textrm{N\'eel}}^x(100)$}
  \psfrag{CNy}{\large ${\mathcal C}_{\textrm{N\'eel}}^y(100)$}
    \psfrag{a}{(a)}
    \psfrag{b}{(b)}
     \psfrag{d1}{$\gamma\!=\!1$}
  \psfrag{d-1}{$\gamma\!=\!-1$}
  \psfrag{jp}{ $J'/J$}
\psfrag{h}{$h_{\rm z}$}
\psfrag{0}{0}
\psfrag{1}{1}
\psfrag{2}{2}
\psfrag{3}{3}
\psfrag{4}{4}
\psfrag{5}{5}
\psfrag{6}{6}
\psfrag{-1}{$-1$}
\psfrag{-3}{$-3$}
\psfrag{-2}{$-2$}
\psfrag{-4}{$-4$}
\psfrag{-5}{$-5$}
\psfrag{-6}{$-6$}
\psfrag{0.1}{$0.1$}
\psfrag{0.2}{$0.2$}
\psfrag{0.3}{$0.3$}
\psfrag{0.4}{$0.4$}
\psfrag{0.0}{$0.0$}
\psfrag{0.5}{$0.5$}
\psfrag{0.6}{$0.6$}
\psfrag{0.7}{$0.7$}
\psfrag{0.00}{$0.00$}
\psfrag{0.05}{$0.05$}
\psfrag{0.10}{$0.10$}
\psfrag{0.15}{$0.15$}
\psfrag{0.20}{$0.20$}
\psfrag{0.25}{$0.25$}
\includegraphics[width=230pt]{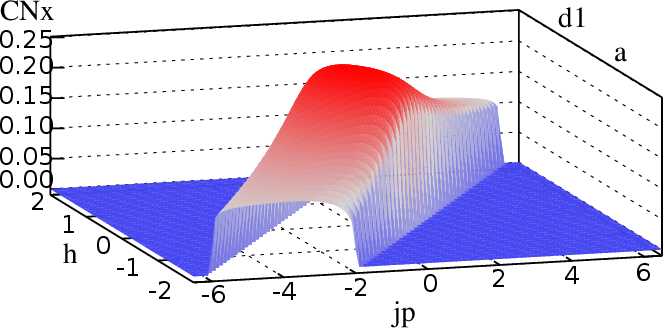}
\includegraphics[width=230pt]{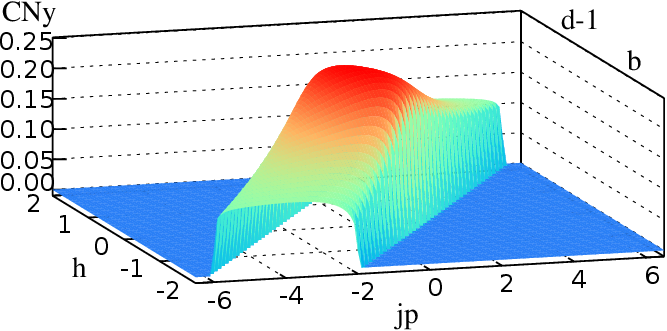}
\caption{Correlation functions, ${\mathcal C}_{\textrm{N\'eel}}^x(n=100)$
  when $\gamma=1$, (a), and ${\mathcal C}_{\textrm{N\'eel}}^y(n=100)$ when
  $\gamma=-1$, (b).}
 \label{Correlation-Neelxy}
\end{figure}
\subsection{Correlation functions}
The analytic expression of spin-spin correlation function can
be obtained easily by casting the spin operators to spinless fermions.
In terms of JW fermions, the correlation function is given
by the products of fermionic operators \cite{LSM,Barouch1,Barouch2}, 
\[{\mathcal C}_{\textrm{N\'eel}}^x(l-m)=\frac{1}{4}\langle B_lA_{l+1}B_{l+1}
\cdots A_{m-1}B_{m-1}A_m\rangle,\]
where $A_j=c_j^\dag+c_j$, and $B_j=c_j^\dag-c_j$.
The above expectation value has been
simplified by using Wick's theorem along with the
fact that $\langle A_lA_m\rangle=\langle B_lB_m\rangle=\delta_{lm}$.
Finally it assumes the from of Toeplitz determinant, 
\be{\mathcal C}_{\textrm{N\'eel}}^x(n)=\frac{1}{4}\left|
\begin{array}{cccc}G_{-1}&G_{-2}&\cdots&G_{-n}\\[0.4em]
  G_{0}&G_{-1}&\cdots&G_{-n+1}\\[0.2em]
  \vdots&\vdots&\ddots&\vdots\\[0.2em]
   G_{n-2}&G_{n-3}&\cdots&G_{-1}\end{array}
\right|,\label{JWCx}\ee
where the PBC is imposed and as a result,
elements of the determinant are nothing but the thermal 
expectation value at temperature $T$, which is given by
\bea
G_{n}&=&\langle B_{n+l}A_{l} \rangle,\nonumber\\[0.4em]
&=&\!\!\frac{1}{\pi}\int_0^\pi\!\! d{k}\,
\frac{\tanh{\left(\frac{E_{k}}{2k_{\rm B}T}\right)}}
{E_{k}}\left(\epsilon_{k}\cos{(kn)}-\Delta_{k}\sin{(kn)}\right).\nonumber
\eea
Similarly, $y$- and $z$-component of correlation functions
in thermal equilibrium are given by
\be{\mathcal C}_{\textrm{N\'eel}}^y(n)=\frac{1}{4}\left|
\begin{array}{cccc}G_{1}&G_{2}&\cdots&G_{n}\\[0.4em]
  G_{0}&G_{1}&\cdots&G_{n-1}\\[0.2em]
  \vdots&\vdots&\ddots&\vdots\\[0.2em]
   G_{-n+2}&G_{-n+3}&\cdots&G_{1}\end{array}
\right|,
\label{JWCy}
\ee
and
\be{\mathcal C}_{\textrm{N\'eel}}^z(n)=\langle S^z\rangle^2-G_{n}G_{-n}/4,\label{JWCz}\ee
where uniform magnetization along $z$-direction, 
\[\langle S^z\rangle=-\frac{1}{2\pi}\int_0^\pi\! d{k}\,\epsilon_{k}\,
\frac{\tanh{\left(\frac{E_{k}}{2k_{\rm B}T}\right)}}
{E_{k}}.\]
However, when $T=0$, the ground state expectation values are given by 
\bea\langle S^z\rangle&=&-\frac{1}{2\pi}\int_0^\pi\! d{k}\,\frac{\epsilon_{k}}{E_{k}},\nonumber \\[0.5em]
G_{n}&=&\frac{1}{\pi}\int_0^\pi\!\! d{k}\,
\frac{\epsilon_{k}\cos{({k}n)}-\Delta_{k}\sin{({k}n)}}
{E_{k}}.\nonumber \eea
Few limiting values can be derived easily. For example,
for $\gamma=0$, ${\mathcal C}_{\textrm{N\'eel}}^x(n)={\mathcal C}_{\textrm{N\'eel}}^y(n)$.
It happens due to the fact that $G_{-n}=G_{n}$, when $\gamma=0$. 
In addition, ${\mathcal C}_{\textrm{N\'eel}}^x(n,\pm \gamma)
={\mathcal C}_{\textrm{N\'eel}}^y(n,\mp \gamma)$. 
When $\gamma\gg 1$, ${\mathcal C}_{\textrm{N\'eel}}^x(n)=0$,
for any values of $n$, since the values of $G_{n}$ along the columns
in the Eqs. \ref{JWCx}, and \ref{JWCy} are becoming the same. 
Which means N\'eel correlation does not survive when the NN
parallel spin cooper-pairing is very strong. 
Variation of Correlation functions, ${\mathcal C}_{\textrm{N\'eel}}^x(n=100)$
when $\gamma=1$, and ${\mathcal C}_{\textrm{N\'eel}}^y(n=100)$ when
$\gamma=-1$, are shown in Fig \ref{Correlation-Neelxy} (a)
and (b), respectively. They look identical reflecting the 
fact that with the sign reversal of $\gamma$,
correlations along $x$ and $y$ directions
are interchangeable. Additionally, correlations
preserve the symmetry, 
${\mathcal C}_{\textrm{N\'eel}}^\beta(n)(-J',-h_{\rm z})=
{\mathcal C}_{\textrm{N\'eel}}^\beta(n)(J',h_{\rm z})$, which
actually corresponds to 
the symmetry of Hamiltonian shown in Eq. \ref{Sym2}.
${\mathcal C}_{\textrm {N\'eel}}^\beta(100)$ has the maximum value
at $h_{\rm z}=0$, which decreases with the increase of $|h_{\rm z}|$.
It means quantum fluctuations reduces with the increase of $|h_{\rm z}|$,
since the effect of competing exchange integrals, $J$ and $J'$
is losing as a consequence.  The LRO found in this system does not
survive at nonzero temperatures, since the values
of ${\mathcal C}_{\textrm{N\'eel}}^x(n)$ and
${\mathcal C}_{\textrm{N\'eel}}^y(n)$ vanish for arbitrary $n$ when  $T>0$.
These results are in accordance with the Mermin-Wagner theorem
which states that continuous symmetries of a
system with short-range interactions
cannot be broken spontaneously when $T>0$, in dimensions two or less\cite{MW}. 

\begin{figure}[h]
\psfrag{a}{\large (a)}
\psfrag{b}{\large (b)}
\psfrag{c}{\large (c)}
\psfrag{d}{\large (d)}
\psfrag{e}{\large (e)}
\psfrag{f}{\large (f)}
\psfrag{g}{\large (g)}
\psfrag{h}{\large (h)}
\psfrag{i}{\large (i)}
\psfrag{y}{\large $g_z$}
\psfrag{z}{\large $g_y$}
\psfrag{m}{$J'<-2J-2h_{\rm z}$}
\psfrag{n}{$-2J\!-\!2h_{\rm z}\!<\!J'\!<\!2J\!-\!2h_{\rm z}$}
\psfrag{p}{$J'>2J-2h_{\rm z}$}
\psfrag{1}{\large $J'<-2J$}
\psfrag{2}{\large $-2J<J'<2J$}
\psfrag{3}{\large $J'>2J$}
\psfrag{7}{$J'<-2J+2h_{\rm z}$}
\psfrag{8}{$-2J\!+\!2h_{\rm z}\!<\!J'\!<\!2J\!+\!2h_{\rm z}$}
\psfrag{9}{$J'>2J+2h_{\rm z}$}
\psfrag{4}{\large $\nu=0$}
\psfrag{5}{\hskip 0.0 cm \large $\nu =1$}
\includegraphics[width=250pt]{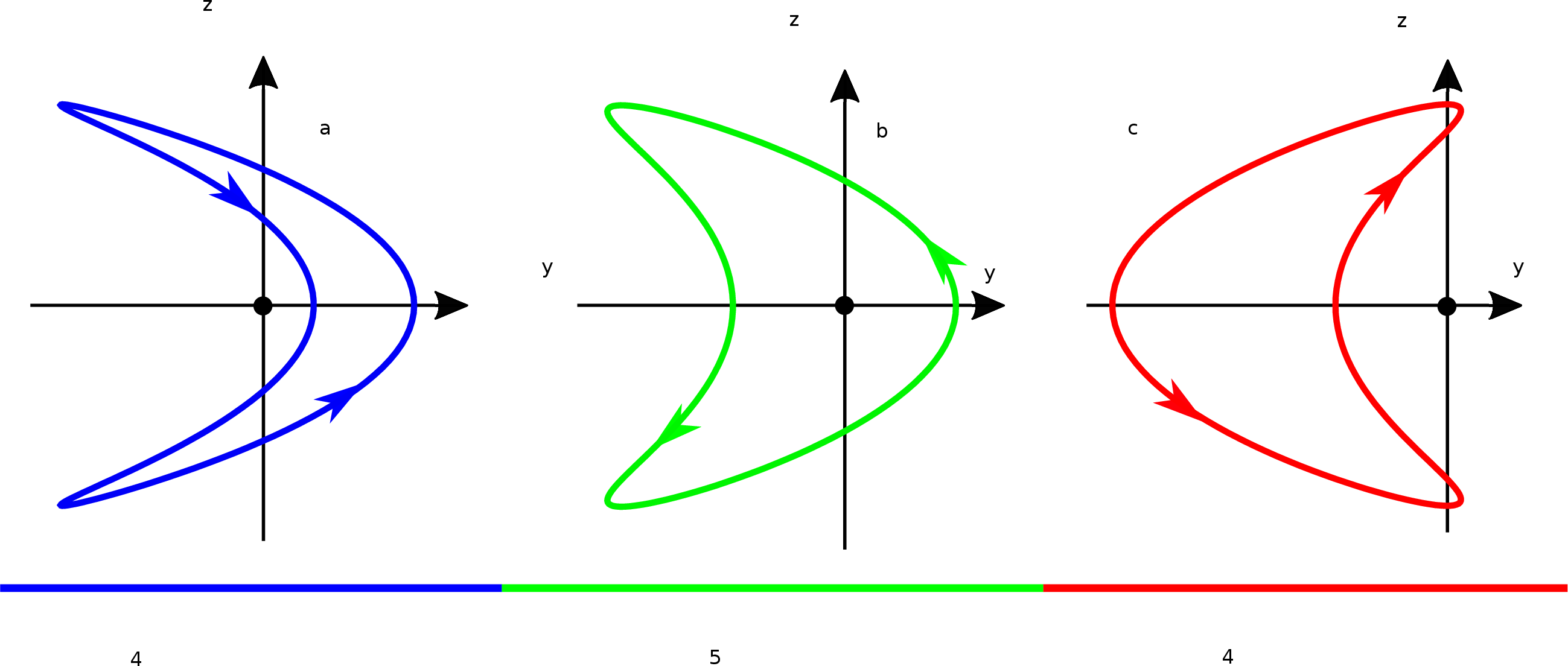}
\vskip .7 cm
\includegraphics[width=250pt]{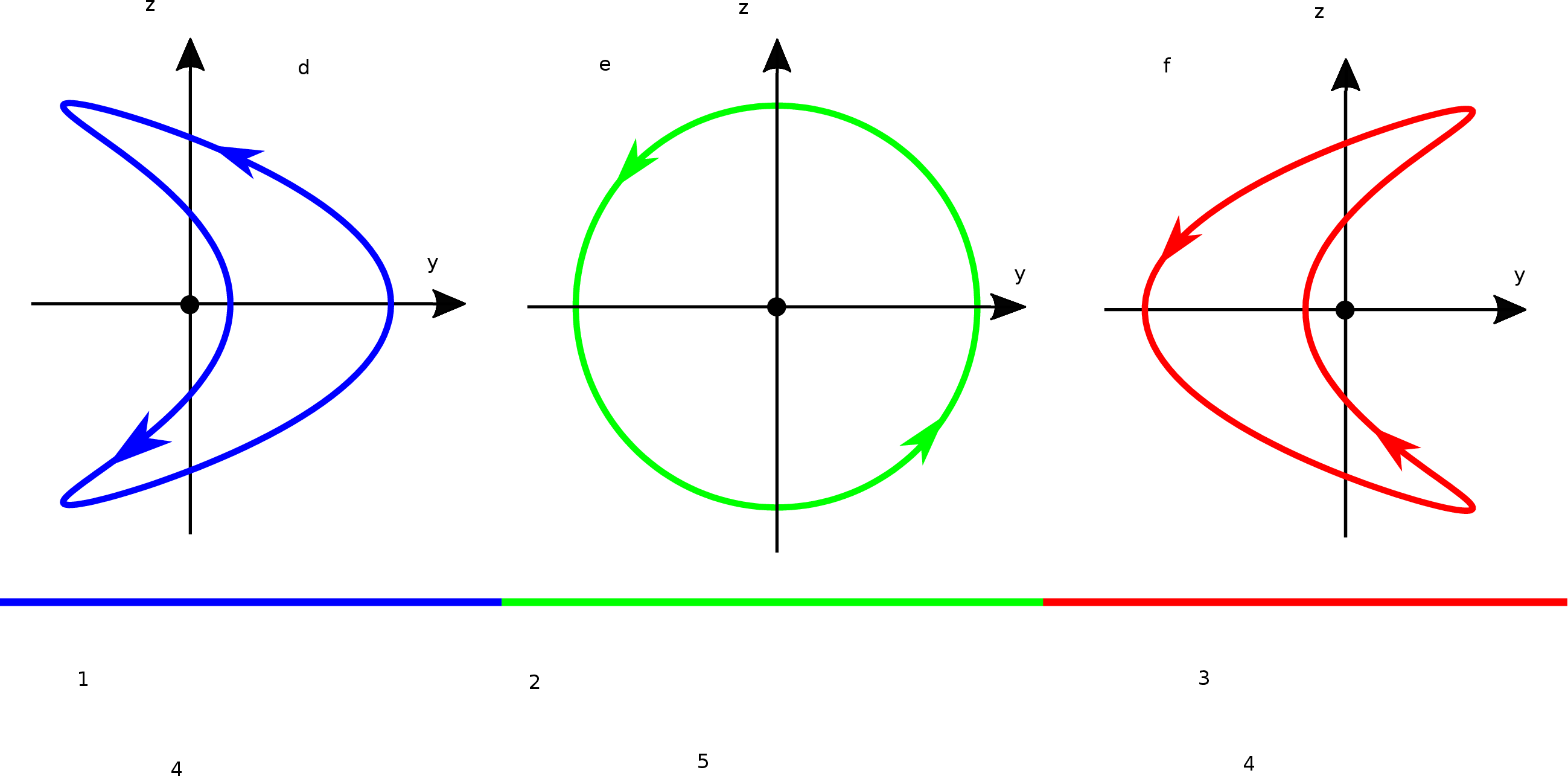}
\vskip .7 cm
\includegraphics[width=250pt]{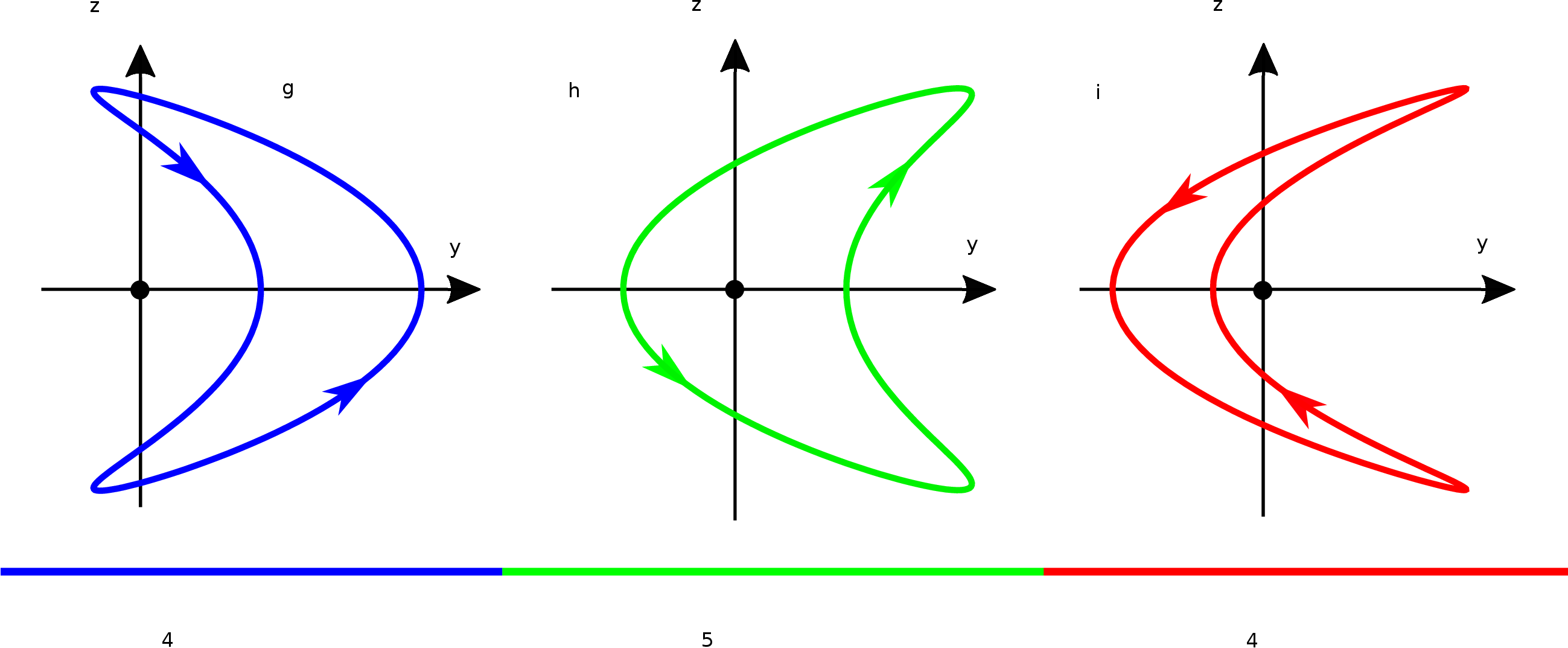}
\caption{Closed contours in the $g_z$-$g_y$ plane for $\gamma=1$.
  Figures (a), (d) and (g) are plotted for $J'/J=-3$, when $h_{\rm z}=-1,0,1$,
  respectively. Similarly, (b), (e), (h) are for $J'/J=0$,  and
  (c), (f), (i) are for $J'/J=3$ when $h_{\rm z}=-1,0,1$,
  respectively. 
     In every case, direction of winding is counter clockwise. For $\gamma=-1$,  
   shape of the contours will be the same but with winding of clockwise direction.}
 \label{winding-123}
\end{figure}
\section{Topological properties of $H$:}
\label{topology}
\subsection{Pfaffian invariant and winding number}
For $\gamma \ne 0$, 
$\mathcal H({k})$ satisfies the time-reversal, particle-hole,
and chiral symmetries as it obeys the following 
relations,
\[\left\{\begin{array}{l}
\mathcal P \mathcal H({k}) \mathcal P^{-1}=-\mathcal H(-{k}),\\ [0.4em]
\mathcal K  \mathcal H({k}) \mathcal K^{-1}=\mathcal H(-{k}),\\ [0.4em]
\mathcal C  \mathcal H({k}) \mathcal C^{-1}=-\mathcal H({k}),\end{array}\right. \]
respectively. Here $\mathcal K$, $\mathcal P=\sigma_x\mathcal K$,
and $\mathcal C=\sigma_x$ are the complex conjugation, particle-hole and
chiral operators, respectively.
Conservation of any two symmetries corresponds to
that of the remaining one as 
they constitute the BDI class.
In order to characterize the topological phase and
locate the phase transition points, 
sign of Pfaffian invariant and value of winding number have been
evaluated. Topological transition is marked by the change of sign
of the Pfaffian at the transition point. Following the
general technique, Hamiltonian $\mathcal H({k})$ is being converted to
a skewsymmetric albeit hermitian matrix by the transformation,
$\tilde H({k})=\mathcal D \mathcal H({k})\mathcal D^\dag$, where
\[\mathcal D=\frac{1}{\sqrt 2}\!
\left[\begin{array}{cc}1 & \;\, i
     \\[0.2em]     1&\;- i
       \end{array} \right]\!.\]
In order to locate the phase transition point, momentum-space 
Pfaffian of the matrix $i\tilde H({k})$ is determined,
which is defined as pf$[i\tilde H({k})]=\sqrt{det\{i\tilde H({k})\}}$\cite{Sau}. 
As the $E_{\rm Gap}$ vanishes exactly at ${k}=\pi$, and ${k}=0$, for
$J'= 2J+2h_{\rm z}$, and $J'= -2J+2h_{\rm z}$, respectively,
Pfaffians, pf$[i\tilde H({k}=\pi)]$, and pf$[i\tilde H({k}=0)]$
are evaluated. These values are given by
\bea {\rm pf}[i\tilde H(\pi)]&=&\frac{1}{2}\left(-J-\frac{J'}{2}+h_{\rm z}\right)\!,\nonumber\\[0.4em]
{\rm pf}[i\tilde H(0)]&=&\frac{1}{2}\left(J-\frac{J'}{2}+h_{\rm z}\right)\!.\nonumber
\eea
Hence the sign of ${\rm pf}[i\tilde H(\pi)]$ and
${\rm pf}[i\tilde H(0)]$ obey the relations:
\bea sign({\rm pf}[i\tilde H(\pi)])&=&\left\{\begin{array}{cc}
-ve, & J'> 2J+2h_{\rm z},\\ [0.4em]
-ve, & -2J+2h_{\rm z}<J'< 2J+2h_{\rm z},\\ [0.4em]
+ve, & J'< -2J+2h_{\rm z},\end{array}\right.\nonumber\\[0.4em]
sign({\rm pf}[i\tilde H(0)])&=&\left\{\begin{array}{cc}
-ve, & J'> 2J+2h_{\rm z},\\ [0.4em]
+ve, & -2J+2h_{\rm z}<J'< 2J+2h_{\rm z},\\ [0.4em]
+ve, & J'< -2J+2h_{\rm z}.\end{array}\right.\nonumber\eea
Now, the Pfaffian invariant,  
$Q=sign({\rm pf}[i\tilde H(\pi)])\times sign({\rm pf}[i\tilde H(0)])$,
is given by\cite{Sau}
\be
Q=\left\{\begin{array}{cc}
+ve, & J'> 2J+2h_{\rm z},\\ [0.4em]
-ve, & -2J+2h_{\rm z}<J'< 2J+2h_{\rm z},\\ [0.4em]
+ve, & J'< -2J+2h_{\rm z}.\end{array}\right.\nonumber
\ee
This result indicates that $Q$ is negative in the
annular region enclosed by the boundary lines
$J'= 2J+2h_{\rm z}$, and $J'= -2J+2h_{\rm z}$, while
$Q$ is positive elsewhere. 
In order to characterize the topology, value of 
bulk topological invariant, {\it i. e.},  winding number ($\nu$) 
has been determined, which is defined as
\[\nu=\frac{1}{2\pi}\oint_C \left(\boldsymbol {\hat g}(k)\times
  \frac{d}{dk}\,\boldsymbol {\hat g}(k)\right)\!d{k},\]
where $\boldsymbol {\hat g}(k)
=\boldsymbol g(k)/|\boldsymbol g(k)|$, and $C$
is a closed curve in the $g_z$-$g_y$ plane. Winding number enumerates
the number of winding around the origin, and at the same time,
it will be accounted as positive when the
curve $C$ is traversed along the counter clockwise direction.
For $\gamma=1$, three different sets of contours,
\{(a), (b), (c)\}, \{(d), (e), (f)\}, and \{(g), (h), (i)\},
as shown in Fig \ref{winding-123}
are drawn for $h_{\rm z}=-1,\,0,\,1$, respectively.
In each triplet set, three different contours are consecutively drawn for
$J'/J=-3,\,0,\,+3$. However, in each diagram direction of the contour
is counter clockwise. So, the value of winding number, $\nu=1$ for
the diagrams (b), (e) and (h), since in each case it encloses the
origin once. Whereas, for the remaining diagrams it is zero as they do not
enclose the origin. On the other hand, for $\gamma=-1$, the shapes of the
contours remain same but the winding is around clockwise direction. 
Hence the nontrivial topological phase is defined by $\nu=-1$, in this case.
When $\gamma>0$ ($\gamma<0$), value of winding number 
for $H$ in the parameter space is given by 
\be
\nu=\left\{\begin{array}{ll}
    1 \,(-1),& -2J+2h_{\rm z}<J'<2J+2h_{\rm z},\\[0.5em]
    0,&J'>2J+2h_{\rm z}\; {\rm and}\;J'<-2J+2h_{\rm z}.
  \end{array}\right. \nonumber
\ee
So, the points, $\gamma=0$, and $J'=2(h_{\rm z}\pm J)$, are actually the 
    bicritical points. 
\begin{figure}[h]
\psfrag{E}{Energy/$J$}
\psfrag{a}{(a)}
\psfrag{c}{(c)}
\psfrag{b}{(b)}
\psfrag{jp}{$J'/J$}
  \includegraphics[width=120pt,angle=-90]{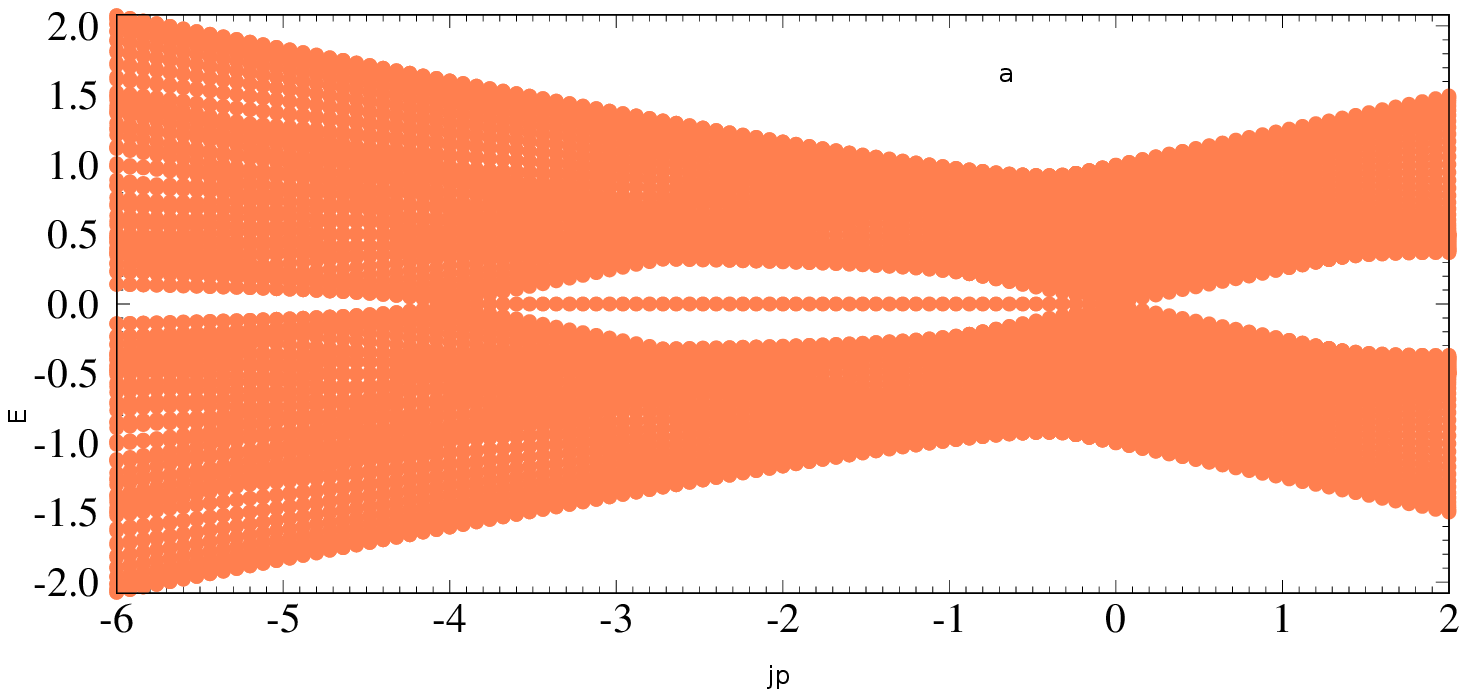}
 \vskip 0.3cm
  \includegraphics[width=120pt,angle=-90]{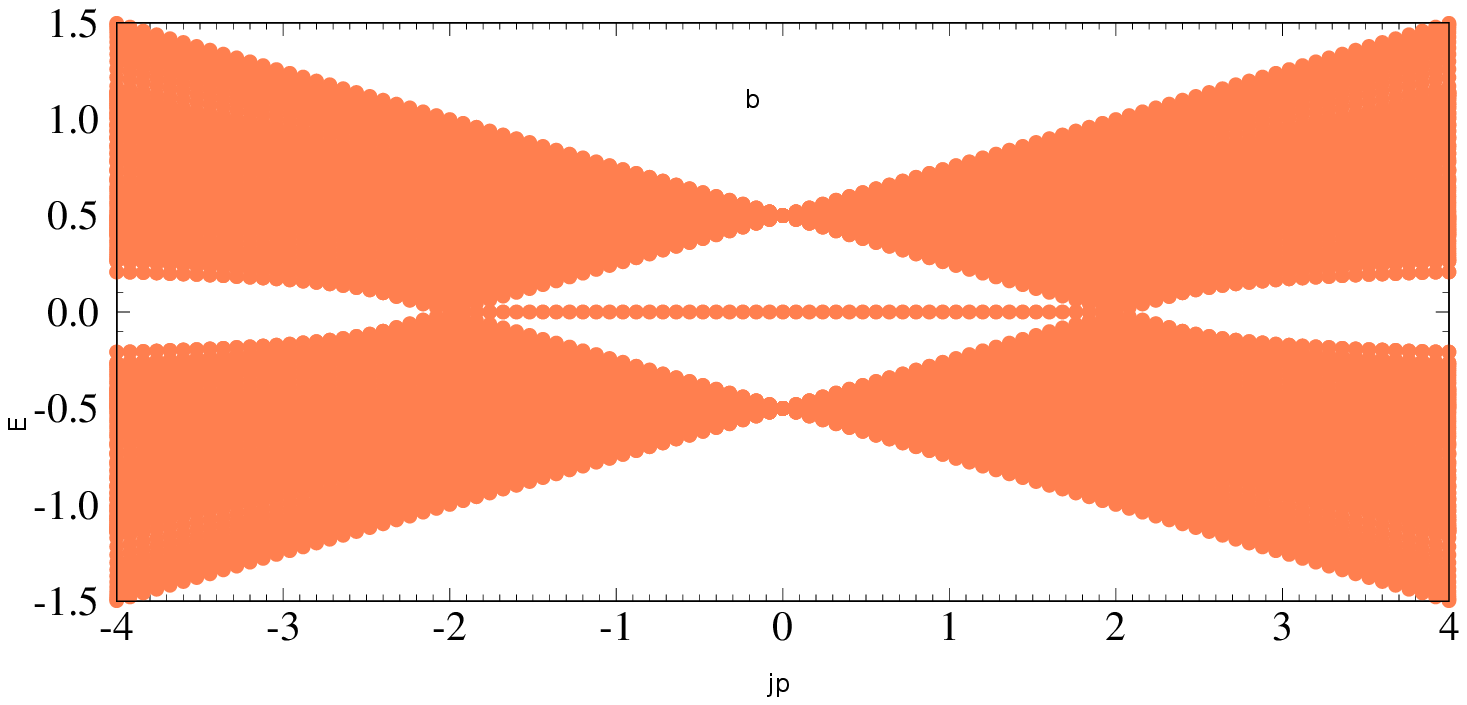}
  \vskip 0.3cm
  \includegraphics[width=120pt,angle=-90]{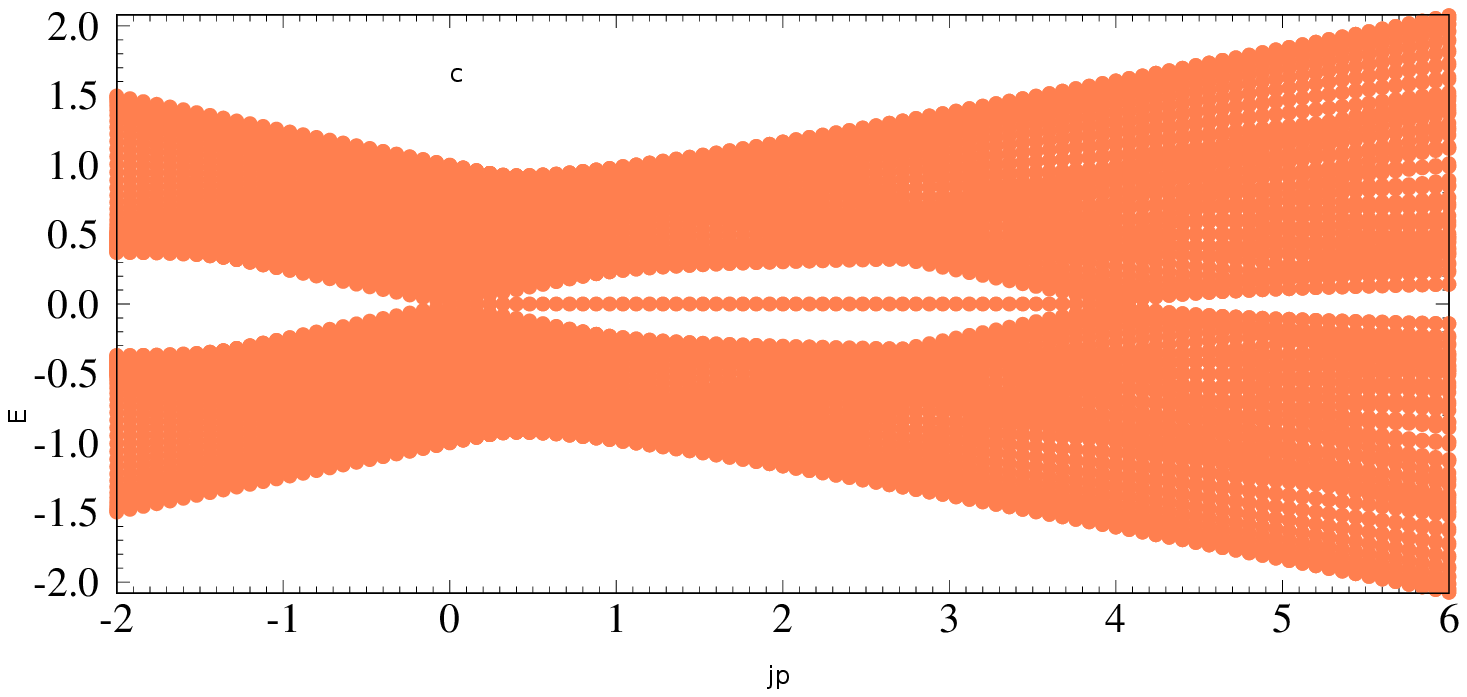}
\caption{Variation of eigen energies with respect to $J'/J$ for $h_{\rm z}=-1$
  (a), $h_{\rm z}=0$ (b), and $h_{\rm z}=1$ (c) when $J=1$, and
  $\gamma=1$. Zero energy edge states
  are present in the topological region.}
\label{Edge-states}
\end{figure}

\begin{figure}[h]
\psfrag{h}{$h_{\rm z}/J$}
\psfrag{A}{\hskip 3.398 cm Antiferromagnet}
\psfrag{t}{\hskip 1.2 cm Trivial}
\psfrag{w1}{\hskip 3 cm $\nu=\pm 1$}
\psfrag{w0}{\hskip 1.5 cm $\nu=0$}
\psfrag{jp}{$J'/J$}
  \includegraphics[width=230pt,angle=0]{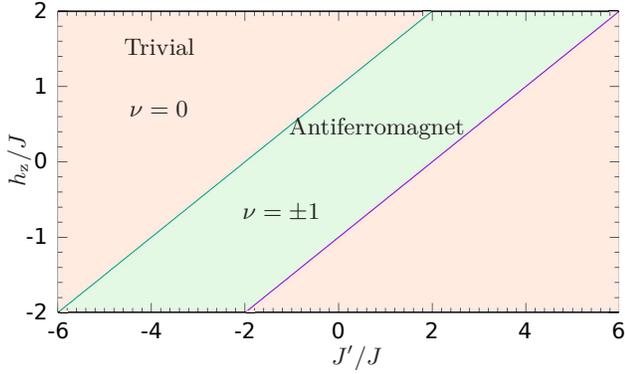}
  \caption{Magnetic and topological phase diagram in the $J'-h_{\rm z}$ space.
    Topological phases of  $\nu= \pm 1$
    is found to coexist with AFM phase with ordering along
    $x$ and $y$ directions, respectively for $J>0$,
    when $\gamma>0$ and $\gamma<0$. So, the points,
    $\gamma=0$, and $J'=2(h_{\rm z}\pm J)$, are the 
    bicritical points. AFM phase will be replaced by FM phase
  if $J<0$.}
\label{Phase-diagram}
\end{figure}
In order to check the bulk-boundary correspondence rule\cite{Hatsugai},
variation of eigen energies for 
$H$ under open boundary condition is shown in Fig. \ref {Edge-states},
with respect to $J'/J$, for $h_{\rm z}=-1$
  (a), $h_{\rm z}=0$ (b), and $h_{\rm z}=1$ (c) when $J=1$, and
  $\gamma=1$. Pair of zero energy edge states
are present in the topologically nontrivial region,
$2(h_{\rm z}-J)<J'<2(J+h_{\rm z})$. Finally a combined phase
diagrams for magnetic and topological phases is shown in
Fig. \ref{Phase-diagram}. 
Faithful coexistence of AFM LRO with nontrivial
topological order is noted in the $J'-h_{\rm z}$ space.
The parallel boundary lines enclosing this region are
denoted by the equations: $J'/2=\pm J+h_{\rm z}$.
Energy gap vanishes over those lines, as well as, 
the system undergoes simultaneous phase transition
of magnetic and topological orders on those lines.
\subsection{Domino model}
\begin{figure}[h]
  \psfrag{g}{$\Upsilon_j^a$}
   \psfrag{h}{$\Upsilon_j^b$}
   \psfrag{a}{(a)}
   \psfrag{b}{(b)}
    \psfrag{c}{(c)}
    \psfrag{d}{(d)}
    \psfrag{e}{(e)}
  \psfrag{j}{$j$}
  \psfrag{1}{$j\!=\!1$}
  \psfrag{2}{2}
  \psfrag{3}{3}
  \psfrag{4}{4}
   \psfrag{N}{$N$}
   \psfrag{dd}{\hskip 0.2 cm $\cdots$}
     \psfrag{N1}{$N\!-\!1$}
\includegraphics[width=230pt]{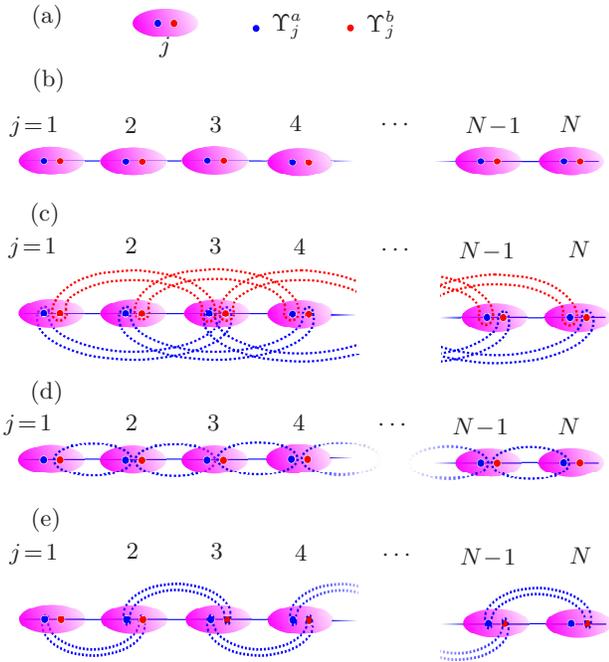}
\caption{Domino model: (a) Spinless fermion at the $j$-th site
  is composed of two Majoranas, 
$\Upsilon_j^a$ and $\Upsilon_j^b$. (b) Lattice composed of Majoranas.
  Majorana pairing in trivial phase (c), and topological phases (d) and (e),
  for the extreme cases.}
     \label{domino-model}
  \end{figure}

The spinless fermions have been converted to Majorana fermions
in order to study the difference of Majorana pairings in
trivial and topological phases.
According to the domino model a normal fermion
at the $j$-th site is made of 
two different Majoranas,
$\Upsilon_j^a$ and $\Upsilon_j^b$, as shown in Fig. \ref{domino-model}
(a), by blue and red dots, respectively
\cite{Beenakker,Alicea,Flensberg,Franchini}. 
Their positions over the chain
is shown in Fig. \ref{domino-model} (b).

So, under the transformation,
\be \Upsilon^a_j=c_j+ c^\dag_j,\;\;
\Upsilon^b_j= i\left(c^\dag_j- c_j\right)\!,\ee
the Hamiltonian in Eq. \ref{hamJW}, for $h_{\rm z}=0$, is written as
\bea
H&=&\frac{i}{4}\sum_{j=1}^{N-1}\left[J(1-\gamma)\Upsilon^a_j\Upsilon^b_{j+1}
-J(1+\gamma)\Upsilon^b_j\Upsilon^a_{j+1}\right]\nonumber\\
&&-\frac{i}{8}J'\sum_{j=1}^{N-2}\left(\Upsilon^a_j\Upsilon^b_{j+2}+\Upsilon^a_{j+2}\Upsilon^b_j\right)\!.
\eea
Three distinct phases can be understood in terms of three different types
of Majorana pairings for three extreme cases.
Among them trivial phase hosts one unique majorana pairing,
while two distinct pairings are found in the nontrivial phase. 
For example, the relations, $J=0$, but $J'\ne 0$, trivial 
corresponds to the trivial phase. 
Total Hamiltonian now reads as
\[H=-i\,\frac{J'}{8}\sum_{j=1}^{N-2}\left(\Upsilon^a_j\Upsilon^b_{j+2}+\Upsilon^a_{j+2}\Upsilon^b_j\right)\!.\]
In this case, a unique 
NNN intercell Majorana pairing has been formed that is 
drawn by red and blue dotted lines as shown in Fig.
\ref{domino-model} (c). 
Obviously, all the Majoranas
participate in the pairing such that no Majorana is left unpaired.
This picture is totally opposite to the trivial phase found in the Kitaev
model, where intracell pairing has been formed\cite{Kitaev}.
So unlike the Kitaev model, NNN Majorana pairing is
found here in the trivial phase. 

Another extreme case leads to a pair of
topological phases which corresponds to
the relations, $J\ne 0$, but $J'= 0$.
Total Hamiltonian for $\gamma=\pm 1$ becomes
\[H= \left\{\begin{array}{lc}-i\,\frac{J}{2}\sum_{j=1}^{N-1}
\Upsilon^b_j\Upsilon^a_{j+1}, & {\rm for}\;\gamma=+1,\\[0.9em]
i\,\frac{J}{2}\sum_{j=1}^{N-1}
  \Upsilon^a_j\Upsilon^b_{j+1},&{\rm for}\;\gamma=-1. \end{array}\right.\]
In this case, two different NN intercell Majorana pairings are permitted.
The first ($\Upsilon_1^a$) and the last ($\Upsilon_N^b$)
Majoranas are left unpaired leading to the appearance of
zero-energy edge states when $\gamma=1$.
This feature is identical to the topological
phase in Kitaev chain which is shown in Fig. \ref{domino-model} (d)
\cite{Kitaev}. Additionally this picture
corresponds to the topological phase with $\nu=1$. 
However, for $\gamma=-1$ the second ($\Upsilon_1^b$) and the
last but second ($\Upsilon_N^a$)
Majoranas are left unpaired which corresponds to the another phase
with $\nu=-1$, as depicted in Fig. \ref{domino-model} (e).  
This particular type of intercell Majorana pairing helps to
construct new fermionic quasiparticle basis in which
the Hamiltonian, $H$ becomes diagonalized in the real space. In order to
accomplish this transformation, new fermionic operators, $\bar c_j$
are constructed by linear superposition of two adjacent Majorana
operators from NN sites\cite{Alicea},
\[\bar c_j=\frac{1}{2}\left(\Upsilon^b_j+\Upsilon^a_{j+1}\right).\]
When $\gamma=1$, in terms of the new fermionic operators,
$H=-J\sum_{j=1}^{N-1}\bar c^\dag_j \bar c_j$,
indicating the cost of creating a new
fermion at any site $j$, is $J$ in the topological phase.
No such fermionic basis can be constructed for the trivial phase
in which $H$ is diagonalized, since two different types
of pairing are superposed in this case.
 
\section{Entanglement: mutual information and quantum discord}
\label{Entanglement}
Entanglement exhibits peculiar behavior at QPTs
and as a result its several measures 
have been employed as a tool to locate the position of critical
points as well as the order of transition\cite{Nielsen,Amico1}.
As a consequence, existence of quantum correlations along with the 
nature of QPTs for the transverse field Ising and anisotropic XY models
have been studied in terms of several measures.
For examples, QPT in transverse Ising model (case {\it 2} of section \ref{model} C)
has been identified easily by obtaining the concurrence along with its
derivatives and other measures\cite{Osterloh,Osborne,Sarandy}.
For transverse field anisotropic XY model
(case {\it 4} of section \ref{model} C),  measures like
mutual information, quantum discord, von Neumann and
Renyi entropies have been employed\cite{Fazio,Korepin1,Korepin2}. 
Correlation can have both classical and quantum origins,
where the total correlation can be measured in terms of
quantum mutual information (MI). For a bipartite
system MI ($I$) has been defined in terms of
von Neumann entropy for the density matrix, $\rho$, 
$S(\rho)=-Tr\{\rho\log_{2}\rho\}$,
which is given by 
\be
I(\rho_{AB})=S(\rho_{A})+S(\rho_{B})-S(\rho_{AB}),
\label{MI}
\ee
where $\rho_{A}$ and $\rho_{B}$ are the reduced density
matrices of the subsystems A and B, while $\rho_{AB}$
is the density matrix of the total system.
For the systems with two-qubit states $\rho_{AB}$
assumes simpler form in the basis
$\{|11\rangle,\,|10\rangle,\,|01\rangle,\,|00\rangle\}$\cite{Ali}. 
In this case, the nonzero elements of two-site
density matrix separated by distance of $n$ units, $\rho_{AB}(n)$, are 
given by $\rho_{11}=u_+$, $\rho_{22}=\rho_{33}=w$, 
$\rho_{23}=\rho_{32}=v_+$, $\rho_{14}=\rho_{41}=v_-$, and
$\rho_{44}=u_-$, where
\be\left\{
\begin{aligned}
&u_\pm=1/4\pm\langle S^z\rangle +{\mathcal C}_{\textrm{N\'eel}}^z(n),\\[0.4em]
&w=1/4-{\mathcal C}_{\textrm{N\'eel}}^z(n),\\[0.4em]
&v_\pm={\mathcal C}_{\textrm{N\'eel}}^x (n)\pm {\mathcal C}_{\textrm{N\'eel}}^y(n).
\end{aligned}\right.
\ee
\begin{figure}[h]
  \psfrag{I}{$I(\rho_{AB})$}
  \psfrag{Q}{$Q(\rho_{AB})$}
    \psfrag{a}{(a)}
    \psfrag{b}{(b)}
     \psfrag{d1}{$\gamma\!=\!1$}
  \psfrag{jp}{ $J'/J$}
\psfrag{h}{$h_{\rm z}$}
\psfrag{0}{0}
\psfrag{1}{1}
\psfrag{2}{2}
\psfrag{3}{3}
\psfrag{4}{4}
\psfrag{5}{5}
\psfrag{6}{6}
\psfrag{-1}{$-1$}
\psfrag{-3}{$-3$}
\psfrag{-2}{$-2$}
\psfrag{-4}{$-4$}
\psfrag{-5}{$-5$}
\psfrag{-6}{$-6$}
\psfrag{0.00}{$0.00$}
\psfrag{0.10}{$0.10$}
\psfrag{0.12}{$0.12$}
\psfrag{0.02}{$0.02$}
\psfrag{0.06}{$0.06$}
\psfrag{0.08}{$0.08$}
\psfrag{0.04}{$0.04$}
\psfrag{0.1}{$0.1$}
\psfrag{0.2}{$0.2$}
\psfrag{0.3}{$0.3$}
\psfrag{0.4}{$0.4$}
\psfrag{0.0}{$0.0$}
\psfrag{0.5}{$0.5$}
\psfrag{0.6}{$0.6$}
\psfrag{0.8}{$0.8$}
\psfrag{1.0}{$1.0$}
\includegraphics[width=230pt]{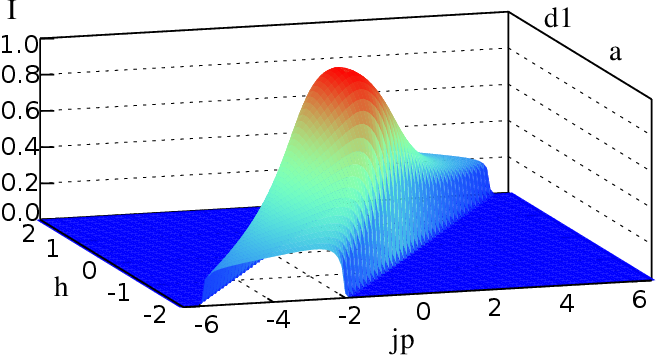}
\includegraphics[width=230pt]{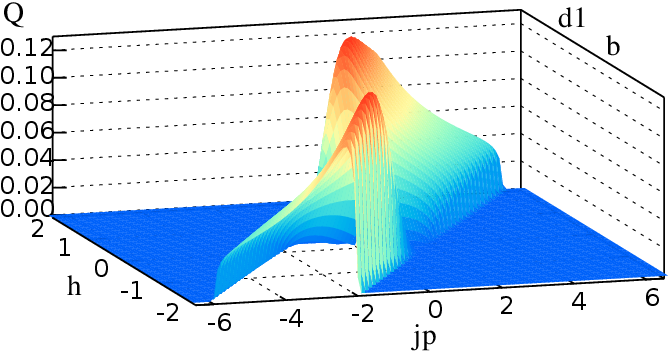}
\caption{Variation of mutual information, $I(\rho_{AB})$ and quantum discord, $Q(\rho_{AB})$
are shown in (a), and (b), respectively, when $\gamma=1$ and $n=100$.}
 \label{MI-QD}
\end{figure}
Matrix elements have been expressed in terms of
magnetization and correlation functions by virtue of the
translational symmetry of the system\cite{Amico2}.
As a result, MI can be written as
\be
I(\rho_{AB})=\sum_{i=1}^4\lambda_i \log_{2}{\lambda_i}-\sum_{\alpha=\pm}\lambda_\alpha \log_{2}{(\lambda_\alpha/2)},
\ee
where, $\lambda_i,\,i=1,2,3,4$, are the eigenvalues of
$\rho_{AB}$, while  $\lambda_\pm=1\pm(u_+-u_-)$.
MI accounts the total correlation, while the quantum
correlation for a bipartite system can be quantified 
by the quantum discord (QD)\cite{Ollivier,Luo,Sarandy}. As long as $\rho_{AB}$
has this simple form, QD ($Q$) can be expressed as
\be
Q(\rho_{AB})=Min\{Q_+,\,Q_-\},
\ee
where
\bea
Q_+\!&=&Q'-\sum_{\alpha=\pm}\left[
  u_\alpha\log_{2}{\left(\frac{u_\alpha}{u_\alpha+w}\right)}
-w\log_{2}{\left(\frac{w}{u_\alpha+w}\right)}\right],\nonumber\\[0.4em]
Q_-\!&=&Q'-\sum_{\alpha=\pm}\Delta_\alpha\log_{2}{\Delta_\alpha},\nonumber
\eea
and
\be\left\{
\begin{aligned}
&Q'=\sum_{i=1}^4\lambda_i \log_{2}{\lambda_i}-\sum_{\alpha=\pm}\frac{\lambda_\alpha}{2} \log_{2}{\left(\frac{\lambda_\alpha}{2}\right)},\\[0.4em]
&\Delta_\pm=\frac{1}{2}(1\pm \Delta),\\[0.4em]
&\Delta=\sqrt{(u_+-u_-)^2+4(|v_+|+|v_-|)^2}.
\end{aligned}\right. \nonumber
\ee
The difference between MI and QD attributes to the
classical correlation (CC), $C(\rho_{AB})=I(\rho_{AB})-Q(\rho_{AB})$.
Actually formulation of QD is accomplished by
maximizing the value of $C(\rho_{AB})$\cite{Henderson}.
Variation of MI and QD are shown in Fig \ref{MI-QD} (a), and (b), respectively, 
when $\gamma=1$ for $n=100$. QD vanishes exactly when $\gamma=\pm 1$, $J'=0$ and 
$h_{\rm z}=0$, due to the fact that at these classical points (Ising limit)
quantum correlation is becoming nil. The nature of this
point is noted before in
case {\it 1} of section \ref{model} C.
The limiting value of MI at this point acquires its maximum value,
$I(\rho_{AB})=1$, which corresponds to the saturation of
classical correlation. 
Apart from that fact, both MI and QD are nonzero in the
region bounded by the lines: $J'/2=\pm J+h_{\rm z}$,
where both magnetic LRO and topological phase coexist. 
Derivatives of $I(\rho_{AB})$ diverge over the lines of
phase transition. 
So, it reveals that these transitions are second order in nature.
Properties of thermal QD and CC for this
system has been studied before\cite{Lin}.  
\section{Discussion}
\label{Discussion}
 Long before Suzuki has proposed a class of generalized XY model
  comprising terms with interaction between more than two spins
 and those are solvable
in terms of spinless fermions\cite{Suzuki1,Suzuki2}. 
Later Titvinidze {\it  et al} has picked up the
simplest Hamiltonian from this class that contains 
three-spin interactions which is termed as
isotropic extended XY model\cite{Japaridze}. 
In this work we have introduced a spin-1/2 1D anisotropic
extended XY model in order to study the interplay of
magnetic and topological phases. In this system, a three-spin
term has been added to the anisotropic XY model, which can be
solved in terms of JW fermions. The existence of long range
magnetic correlations has been confirmed numerically. 
Both magnetic and topological properties
have been studied extensively. System exhibits AFM LRO along two
orthogonal directions, say $x$ and $y$ directions at the
same location in the parameter space, if the anisotropic parameter
$\gamma$ picks up $+ve$ and $-ve$ signs, respectively.
Topological superconducting phases have been characterized
by evaluating the Pfaffian invariant, winding number, and zero-energy 
edge states. System hosts topological phases with
$\nu= \pm 1$ in such a manner that the
coexistence of magnetic and topological
phases in the parameter space along with their  
correspondence is observed, as found before
in the Kitaev model\cite{Kitaev}. 
Additionally, entanglement called MI is also
nonzero in the entire region where this coexistence
is found. While quantum discord is nonzero over the
same region except the Ising point.
These QPTs are of second order since the derivatives of
MI are diverging at the transition points. 
In contrast, QPTs in this model
cannot be detected in terms of entanglement measurements,
like concurrence and fidelity. 

Magnetic and topological properties of this model
at various extreme limits are investigated. It shows that
the anisotropic parameter $\gamma$ plays the crucial role
behind the origin of magnetic LRO as well as the nontrivial topological
phases. Finally, the trivial and topological phases are explained
in terms of different types of Majorana pairings. Two different NN pairings
have appeared in the topological phases for the two extreme cases
defined by $J'=0$, but $\gamma=\pm 1$. On the other hand,
an unique NNN Majorana pairing is found in the other extreme case, 
$J=0$, which corresponds to the trivial phase. In contrast, only NN Majorana
pairing is found in the Kitaev model for both trivial and
topological phases\cite{Kitaev}.

Those phases as well their one-to-one correspondence can be destroyed
by applying transverse magnetic field, beyond $h_{\rm z}/J\ge 1$, 
in the absence of three-spin term.
However, in the presence of three-spin term,
coexistence of those phases
as well their one-to-one correspondence would never be
broken or eliminated by applying the magnetic field of
finite strength. So, in contrast, this model exhibits magnetic LRO
in the magnetic field of any value.
Location of those phases in the parameter
spaces is given by $2(h_{\rm z}-J)<J'<2(J+h_{\rm z})$.
Position of bicritical points are given by
$\gamma=0$, and $J'=2(h_{\rm z}\pm J)$.
On the other hand, QPT can occur even in the absence
of the magnetic field. It occurs due to the fact that magnetic fields 
always tend to destroy the LRO favoured by the 
exchange integrals $J$ and $J'$. However, in this model
$J$ and $J'$ also compete with each other. So, QPT might occur
due to the competing effect of $J$ and $J'$ alone.
In this sense, the three-spin
terms induces exotic magnetic phases which were not
explored before. Magnetic field always opposes the LRO 
in systems of any dimension
\cite{Barouch1,Barouch2,Barouch3,Mikeska,Franchini,Kurmann1,Kurmann2,Muller,Ghosh1,Ghosh2}.
On the other hand, there are several instances where 
magnetic field is indispensable for the emergence of
topological phases specially in the magnetic systems
\cite{Owerre1,Joshi,Moumita1,Sil1,Moumita2,Owerre2,Moumita3,Owerre3,Bhowmick,Moumita4}.
However, in this work, effect of magnetic field both on the magnetic
and topological phases are investigated and their interplay 
has been explored. Coexistence of magnetic and superconducting phases
has been observed in a number of experiments, however,
faithful coexistence of magnetic and topological superconducting orders
is yet to be demonstrated. QPTs found in this model could have been
determined by three different ways: (i) noting
the variation of spin-spin correlation
functions in the ground state, (ii) finding the value of
topological invariant in the single particle excitation, and
(iii) studying the variation of mutual information and quantum discord
in the many-particle ground state.

 
Topological phase has been experimentally observed
by realizing the tight-binding model on a lattice of finite number of sites
in a number of ways. And in every
case nontrivial phase is determined by examining the 
presence of symmetry protected edge states along with the band gap.
Using a system of bosonic Rydberg atoms coupled
with dipolar exchange interaction, properties of a 
tight-binding model with arbitrary hopping could be tested.
In this technique, the non-interacting fermionic system has been prepared 
by interacting hard-core bosons where the occupation number per site
does not exceed by one due to the hard-core condition. In this method, zero-energy 
edge states of topological SSH model with 14 sites trapped with
${}^{87}$Rb atoms has been successfully demonstrated\cite{Browaeys}.
Zero-energy edge modes of the same model with ultracold
${}^{84}$Sr atoms has been observed recently in another attempt\cite{Kanungo}.
Topological edge modes and phase transition of the SSH model has been demonstrated 
by a mechanical system composed of mass and elastic string\cite{Merlo}.
Similarly, topological properties of the SSH model have been verified
in several 1D artificial lattices, say in phononic crystal\cite{Wen},
photonic crystal\cite{Poli}, plasmonic crystals\cite{Giannini} and
resonant metamaterials\cite{Zhu}.
The system under this investigation includes NNN hopping besides the NN term.
In order to demonstrate the topological phase of this system,
artificial 1D crystals incorporating NN and NNN interactions
can be generated using any methods as described above.
  \section{ACKNOWLEDGMENTS}
  RKM acknowledges the DST/INSPIRE Fellowship/2019/IF190085.
  \section{Data availability statement}
  All data that support the findings of this study are included within the article.
  \section{Conflict of interest}
  Authors declare that they have no conflict of interest.
   \appendix
\section{ENERGY EIGENVALUES AND EIGENSTATES OF THE FOUR-SITE HAMILTONIAN, $H$}
\label{appendix:eigensystem}
Expressions of 16 eigenvalues of the four-site Hamiltonian, $H$are
as follows.
\vskip 0.2cm 
$E_1=-J-\sqrt{J^2\gamma^2+ (h_{z}+J'/2)^2}$,

$E_2=-J+\sqrt{J^2\gamma^2+ (h_{ z}+J'/2)^2}$, 

 $E_3=J-\sqrt{J^2\gamma^2+ (h_{ z}+J'/2)^2}$,

 $E_4=J+\sqrt{J^2\gamma^2+ (h_{ z}+J'/2)^2}$,

 $E_5=-\sqrt{J^2(1\!+\!\gamma^2)+2h_{ z}^2\!+\!\sqrt{(J^2(1\!+\!\gamma^2)+2h_{ z}^2)^2-8J^2h_{ z}^2}}$,

 $E_6=\sqrt{J^2(1\!+\!\gamma^2)+2h_{ z}^2\!+\!\sqrt{(J^2(1\!+\!\gamma^2)+2h_{ z}^2)^2-8J^2h_{ z}^2}}$,

$E_7=-\sqrt{J^2(1\!+\!\gamma^2)+2h_{ z}^2\!-\!\sqrt{(J^2(1\!+\!\gamma^2)+2h_{z}^2)^2-8J^2h_{ z}^2}}$,

$E_8=\sqrt{J^2(1\!+\!\gamma^2)+2h_{ z}^2\!-\!\sqrt{(J^2(1\!+\!\gamma^2)+2h_{ z}^2)^2-8J^2h_{ z}^2}}$,

$E_9=E_{11}=E_{12}=E_{13}=0$,

$E_{13}=E_{14}=-h_{ z}+J'/2$,

$E_{15}=E_{16}=h_{z}-J'/2$.

\vskip 0.2cm 
To express the eigenfunctions in a compact form,
following notations are used.
\vskip 0.2cm 
$\chi_n^2=T^{n-1}\ket{2} (n=1)$,$\ket{2}=\ket{\uparrow\uparrow\uparrow\uparrow}$,

$\chi_n^1=T^{n-1}\ket{1} (n=1,2,3,4)$,$\ket{1}=\ket{\downarrow\uparrow\uparrow\uparrow}$,

$\chi_{n,1}^0=T^{n-1}\ket{0} (n=1,2,3,4)$,$\ket{0}=\ket{\uparrow\uparrow\downarrow\downarrow}$,

$\chi_{n,2}^0=T^{n-1}\ket{0} (n=1,2)$,$\ket{0}=\ket{\uparrow\downarrow\uparrow\downarrow}$,

$\chi_{n}^{-1}=T^{n-1}\ket{-1} (n=1,2,3,4)$,$\ket{-1}=\ket{\uparrow\downarrow\downarrow\downarrow}$,

$\chi_{n}^{-2}=T^{n-1}\ket{-2} (n=1)$,$\ket{-2}=\ket{\downarrow\downarrow\downarrow\downarrow}$.

Here the operator $T$ behaves as $T\ket{pqrs}=\ket{spqr}$. 
The normalized eigenstates now can be expressed as follows.

\vskip 0.2cm 
$\psi_i=\frac{1}{2\sqrt{\zeta_i^2+1}}\sum_{n=1}^4(-1)^n(\chi_n^{-1}+\zeta_i\chi_n^1)$, $i=1,2$,

$\psi_k=\frac{1}{2\sqrt{\zeta_k^2+1}}\sum_{n=1}^4(-\chi_n^{-1}+\zeta_k\chi_n^1)$, $k=3,4$,

$\psi_l=\frac{1}{2N_l}[(E_l^2-4h_z^2)(E_l\sum_{n=1}^4\chi_{n,1}^0+2J\sum_{n=1}^2\chi_{n,2}^0)$

 $ +2\gamma J E_l\{(E_l-2h_z)\chi_1^{-2}+(E_l+2h_z)\chi_1^{2}\}]$, $l=5,6,7,8$,

$\psi_m=\frac{1}{\sqrt{2}}(\chi_n^0-\chi_{n+1}^0)$, $n=1,2,3$ and $m=9,10,11$,

$\psi_{12}=\frac{1}{\sqrt{2}}\sum_{n=1}^{2}(-1)^{n(n-1)/2}\chi_{n,2}^0$,

$\psi_{13}=\frac{1}{2}\sum_{n=1}^{4}(-1)^{n(n-1)/2}\chi_n^{-1}$,

$\psi_{14}=\frac{1}{2}\sum_{n=1}^{4}(-1)^{n(n+1)/2-1}\chi_n^{-1}$,

$\psi_{15}=\frac{1}{2}\sum_{n=1}^{4}(-1)^{n(n+1)/2}\chi_n^{1}$,

$\psi_{16}=\frac{1}{2}\sum_{n=1}^{4}(-1)^{n(n-1)/2}\chi_n^{1}$,

with $\zeta_i=\frac{\gamma J}{h_z-J+J'/2-E_i}$,

$\zeta_k=\frac{\gamma J}{h_z+J+J'/2-E_k}$ and 

$N_l=\sqrt{(\sqrt{2}\gamma J E_l)^2\!+\!(E_l^2\!+\!2J^2)(E_l^2\!-\!4h_z^2)^2\!+\!(2\sqrt{2}\gamma J E_l h_z)^2}$.



\begin{thebibliography}{99}
  \bibitem{Sachdev}S Sachdev, Quantum Phase Transitions, Cambridge University
    Press, Cambridge, (1999).
    \bibitem{BKC1}B K Chakrabarti, A Dutta, P Sen, Quantum Ising Phases and
  Transitions in Transverse Ising Models, Springer, Berlin (1995).
\bibitem{BKC3}A Dutta, G Aeppli, B K Chakrabarti, U Divakaran,
  T F Rosenbaum and D Sen, Quantum Phase Transitions in Transverse
  Field Spin Models, Cambridge University Press, Cambridge, (2015)
\bibitem{Franchini}  F Franchini, An introduction to integrable
  techniques for one-dimensional quantum systems, Springer, Heidelberg (2017).
\bibitem{LSM} E. Lieb, T. Schultz, and D. Mattis, Ann. Phys. {\bf 16},
  407 (1961)
\bibitem{Katsura} S. Katsura, Phys. Rev. {\bf 127}, 1508 (1962).
\bibitem{Barouch1} E. Barouch, B. M. McCoy and M Dresden, Phys. Rev. A {\bf 2}, 1075 (1970).  
\bibitem{Barouch2} E. Barouch and B. M. McCoy, Phys. Rev. A {\bf 3}, 786 (1971).
\bibitem{Barouch3} E. Barouch and B. M. McCoy, Phys. Rev. A {\bf 3}, 2137 (1971).
\bibitem{Pfeuty} P. Pfeuty, Ann. Phys. {\bf 57}, 79 (1970).
\bibitem{JW}P. Jordan, E. Wigner, Z. Phys. 47, 631 (1928)
\bibitem{Coldea} R Coldea, D A Tennant, E M Wheeler, E Wawrzynska,
  D Prabhakaran, M Telling, K Habicht, P Smeibidl, K Kiefer, Science, {\bf 327}, 177 (2010).
\bibitem{Blinc} R. Blinc, J. Phys. Chem. Solids {\bf 13}, 204 (1960).
\bibitem{Stinchcombe}R B Stinchcombe, J . Phys. C: Solid State Phys., {\bf  6}, 2459 (1973).
\bibitem{Osterloh}Osterloh A, Amico L, Falci G and Fazio R, Nature {\bf 416} 608 (2002)
\bibitem{Osborne} T. J. Osborne, and M. A. Nielsen, Phys. Rev. A {\bf 66}, 032110 (2002).  
\bibitem{Korepin1} F Franchini, A R Its, B-Q Jin and V E Korepin,
  J. Phys. A: Math. Theor. {\bf 40} 8467 (2007)
    
\bibitem{Korepin2}F Franchini, A R Its and V E Korepin,
  J. Phys. A: Math. Theor. {\bf 41} 025302 (2008)
\bibitem{Franchini1} F. Franchini, A. R. Its,  V. E. Korepin and  
L. A. Takhtajan, Quantum Inf Process {\bf 10}, 325-341 (2011)
\bibitem{Nielsen} Nielsen M and Chuang I, Quantum Computation and
  Quantum Communication, Cambridge Univ. Press, Cambridge, (2000).
\bibitem{Klitzing} Klitzing K. V., Rev. Mod. Phys. {\bf 58}, 519 (1986). 
\bibitem{TKNN}  Thouless D. J., Kohomoto M., Nightingale P. and den Nijs M.,  
Phys. Rev. Lett. {\bf 49}, 405 (1982).
\bibitem{Haldane} Haldane F. D. M., Phys. Rev. Lett. {\bf 61}, 2015 (1988).
  \bibitem{SSH1} W. Su, J. Schrieffer and A. J. Heeger, 
Phys. Rev. Lett. {\bf 42}, 1698 (1979).
\bibitem{SSH2}  Heeger A. J., Kivelson S., Schrieffer J. R. and  Su W. -P., 
  Rev. Mod. Phys. {\bf 60}, 781 (1988).
  \bibitem{Rakesh}R K Malakar and A K Ghosh, J . Phys. Condens. Matter, {\bf  35}, 335401 (2023).
\bibitem{Hatsugai}Y. Hatsugai, Phys. Rev. Lett. {\bf 71}, 3697 (1993).
\bibitem{Kitaev} A. Y. Kitaev, Phys. -Usp. {\bf 44}, 131, (2001)
\bibitem{Binder1} K. Binder, Phys. Rev. Lett. {\bf 47}, 693 (1981).
\bibitem{Binder2}K. Binder, Z. Phys. B {\bf 43}, 119 (1981).
\bibitem{Ising} E. Ising, Z. Phyzik {\bf 31}, 253 (1925).
\bibitem{Kurmann1}J. Kurmann, G. M\"uller, H. Thomas, M. W. Puga and H. Beck,
  J. Appl. Phys. {\bf 52}, 1968-1970 (1981).
\bibitem{Kurmann2} J. Kurmann, H. Thomas, G.M\"uller, Physica A {\bf 112}, 235 (1982).  
\bibitem{Muller}G. M\"uller and R. E. Shrock, Phys. Rev. B {\bf 32}, 5845 (1985).
\bibitem{Japaridze} I. Titvinidze and G. I. Japaridze,
  Eur. Phys. J. B {\bf 32}, 383-393 (2003)
\bibitem{MW} Mermin N D and Wagner H,  Phys. Rev. Lett. {\bf 17} 1133-6 (1966) 
\bibitem{Sau} J. D. Sau and S Tewari, Phys. Rev. B {\bf 88}, 054503 (2013).
\bibitem{Beenakker} C.W.J. Beenakker, Annu. Rev. Condens. Matter Phys. {\bf 4}, 113 (2013).
\bibitem{Flensberg}M Leijnse and K Flensberg, Semicond. Sci. Technol. {\bf 27}, 124003  (2012).
\bibitem{Alicea}J Alicea, Rep. Prog. Phys. {\bf 75}, 076501 (2012).
\bibitem{Amico1}  L. Amico, R. Fazio, A. Osterloh, V. Vedral, Rev. Mod.
Phys. {\bf 80}, 517 (2008)
\bibitem{Ali} M. Ali, A. R. P. Rau, G. Alber, Phys. Rev. A {\bf 81}, 042105
(2010)  
\bibitem{Amico2}L Amico, A Osterloh, F Plastina, R Fazio
and G M Palma, Phys. Rev. {\bf A} 69, 022304 (2004)  
\bibitem{Ollivier} H. Ollivier, W. H. Zurek,
  Phys. Rev. Lett. {\bf 88}, 017901 (2001)
\bibitem{Luo}S Luo, Phys. Rev. {\bf A} 77, 042303 (2008) 
\bibitem{Sarandy} M. S. Sarandy, Phys. Rev. {\bf A} 80, 022108 (2009)
\bibitem{Lin} Y-C Li and H-Q Lin, Phys. Rev. {\bf A} 83, 052323 (2011) 
\bibitem{Suzuki1} M Suzuki, Phys. Lett. A 34, 94 (1971) 
\bibitem{Suzuki2} M Suzuki, Prog. Theor. Phys. 46, 1337 (1971)  
\bibitem{Henderson} L. Henderson, V. Vedral, J. Phys. A Math. Gen. {\bf 34}, 6899
(2001)
\bibitem{Fazio}A. Osterloh, L. Amico, G. Falci and R. Fazio, Nature {\bf 416}, 11 (2002). 
\bibitem{Mikeska}H -J Mikeska, A Ghosh and A K Kolezhuk, Phys. Rev. Lett. {\bf 93} 217204 (2004).
\bibitem{Ghosh1} A Ghosh, J. Phys: Condens. Matter {\bf 13} 5205 (2001).
\bibitem{Ghosh2} A K Ghosh, Phys. Rev. B {\bf 80} 214418 (2009).
\bibitem{Owerre1} Owerre S A, J. Appl. Phys. {\bf 120}, 043903 (2016).
\bibitem{Moumita1} M Deb and A K Ghosh, J. Phys.: Condens. Matter {\bf 31},
  345601 (2019).
\bibitem{Sil1} Sil A. and Ghosh A. K., J. Phys.: Condens. Matter {\bf 32}, 205601 (2020).    
\bibitem{Joshi} D G Joshi, Phys. Rev. B {\bf 98}, 060405(R) (2018).
\bibitem{Moumita2} M Deb and A K Ghosh, J. Magn. Magn. Mater. {\bf 533}, 167968 (2021).
\bibitem{Owerre2} Owerre S A, J. Phys.: Condens. Matter {\bf 29}, 03LT01 (2016). 
\bibitem{Moumita3} M Deb and A K Ghosh, Eur. Phys. J. B {\bf 93}, 145 (2020).
\bibitem{Owerre3} Owerre S A, J. Phys.: Condens. Matter {\bf 29}, 185801 (2017).
\bibitem{Bhowmick}D Bhowmick and P Sengupta, Phys. Rev. B {\bf 101}, 214403 (2020).
\bibitem{Moumita4} M Deb and A K Ghosh, J. Phys.: Condens. Matter {\bf 32}, 365601 (2020).
\bibitem{Browaeys}S de L\'es\'eleuc, V Lienhard, P Scholl, D Barredo,
  S Weber, N Lang, H P B\"uchler,T Lahaye and A Browaeys, Science {\bf 365}, 775-780 (2019)
\bibitem{Kanungo} S. K. Kanungo, J. D. Whalen, Y. Lu, M. Yuan, K. R. A. Hazzard and T. C. Killian,
  Nat. Commun.  {\bf 13}, 972 (2022)
\bibitem{Merlo} L Thatcher, P Fairfield, L Merlo-Ram\'irez and J M Merlo,  Phys. Scr. {\bf 97}, 035702 (2022)
\bibitem{Wen}X Li, Y Meng, X Wu, S Yan, Y Huang, S Wang, W Wen,
  Appl. Phys. Lett. {\bf 113}, 203501 (2018)
\bibitem{Poli} C Poli, M Bellec, U Kuhl, F Mortessagne and H Schomerus,
  Nat. Commun. {\bf 6} 6710 (2015)
\bibitem{Giannini}S. R. Pocock, X Xiao, P. A. Huidobro, and V Giannini, ACS Photonics  {\bf 5}, 2271-2279 (2018)
\bibitem{Zhu}W Zhu,1 Y Ding, J Ren, Y Sun, Y Li, H Jiang, and H Chen,
  Phys. Rev. B {\bf 97}, 195307 (2018).
\end{thebibliography}
\end{document}